\documentclass[preprint]{emulateapj}

\usepackage{graphicx}
\newcommand{\um}{$\mu$m}
\newcommand{\brgamma}{Br$\gamma$}
\newcommand{\kms}{km\thinspace s$^{-1}$}

\def\degr{\hbox{$^\circ$}}
\def\arcmin{\hbox{$^\prime$}}
\def\arcsec{\hbox{$^{\prime\prime}$}}
\def\utw{\smash{\rlap{\lower5pt\hbox{$\sim$}}}}
\def\udtw{\smash{\rlap{\lower6pt\hbox{$\approx$}}}}

\def\fsa{\hbox{$^{\rm s}$}}

\def\farcsa{\hbox{$^{\prime\prime}$}}

\def\Msun{\hbox{\it M$_\odot$}}

\def\K{\hbox{\it K}}

\newcommand{\Ks}{{\it K$_{\rm S}$}}

\newcommand{\Aks}{{\it A$_{\rm K_{\rm s}}$}}
\newcommand{\Ak}{{\it A$_{\rm K}$}}

\def\simgr{\mathrel{\hbox{\rlap{\hbox{\lower4pt\hbox{$\sim$}}}\hbox{$>$}}}}

\def\cf{\hbox{\it C$_f$}}

\shorttitle{The GLIMPSE9 stellar cluster.}
\shortauthors{Messineo et al.}

\begin{document}


\title{HST/NICMOS observations of the GLIMPSE9 stellar cluster.}


\author{Maria~Messineo\altaffilmark{1,6,7}, 
        Donald~F.~Figer\altaffilmark{1}, 
	Ben~Davies\altaffilmark{1,2,6},
	R.P.~Kudritzki\altaffilmark{3},
        R. Michael Rich\altaffilmark{4}, 
	John MacKenty\altaffilmark{5,6},
	Christine Trombley\altaffilmark{1}}

\email{mmessine@rssd.esa.int}

\altaffiltext{1}{Chester F. Carlson Center for Imaging Science, Rochester Institute
   of Technology, 54 Lomb Memorial Drive, Rochester, NY 14623-5604, United
   States.}

\altaffiltext{2}{School of Physics \& Astronomy, University of Leeds, Woodhouse Lane, Leeds
LS2 9JT, UK.}

\altaffiltext{3}{Institute for Astronomy, University of Hawaii, 2680 
Woodlawn Drive, Honolulu, HI 96822}

\altaffiltext{4}{Physics and Astronomy Building, 430 Portola Plaza, Box 951547, Department of Physics and Astronomy, University of California, Los Angeles, CA 90095-1547.}

\altaffiltext{5}{Space Telescope Science Institute, 3700 San Martin Drive, Baltimore, MD 21218.}

\altaffiltext{6}{Visiting Astronomer, Kitt Peak National Observatory, National Optical Astronomy Observatory, 
which is operated by the Association of Universities for Research in Astronomy (AURA) under cooperative agreement with 
the National Science Foundation.}

\altaffiltext{7}{The Astrophysics and Fundamental Physics Missions Division, Research and
Scientific Suppport Department, Directorate of Science and Robotic Exploration,
ESTEC, Postbus 299, 2200 AG Noordwijk, the Netherlands.
}


%

\begin{abstract}   
We present HST/NICMOS photometry, and low-resolution K-band spectra of the 
GLIMPSE9 stellar cluster.
The newly obtained color-magnitude diagram shows a cluster sequence  with
$H$-\Ks\ $=\sim 1$  mag, indicating an interstellar extinction \Aks$=1.6\pm0.2$
mag. The spectra of the three brightest stars show deep CO band-heads, which
indicate red supergiants with spectral type M1-M2. Two 09-B2 supergiants are
also identified, which yield  a spectrophotometric distance of $4.2\pm0.4$ kpc.
Presuming that the population is coeval, we derive an age between 15 and 27 Myr,
and a  total cluster mass of $1600\pm400$ \Msun, integrated down to 1 \Msun. In
the vicinity of GLIMPSE9 are several HII regions and SNRs, all of which
(including GLIMPSE 9) are probably associated with a giant molecular cloud (GMC)
in the inner galaxy. GLIMPSE9 probably represents one episode of massive star
formation  in this GMC. We have identified several other candidate   stellar
clusters of the same complex. 

\end{abstract}




\section{Introduction}   
An understanding of the mechanisms of  formation and evolution  of  massive
stars is of broad astronomical interest.  Through mass-loss and supernova explosions, massive stars
return a significant fraction of their masses to the interstellar medium (ISM),  thereby chemically
enriching and shaping the ISM.  Being very luminous, they can be identified in external galaxies, 
providing spectrophotometric distance. Observational constraints on the formation and evolution of
massive  stars are, however, difficult to obtain due to  the  rarity of these objects and  their
location in the Galactic plane, where interstellar extinction  can hamper their detection.  

Massive stars can be identified by their ionizing radiation, which creates
easily identifiable HII regions. Furthermore,
since the majority of massive stars are born in clusters \citep{lada03},  they
are also identified by locating  young massive  stellar clusters.   Over the
past decade, infrared and radio observations of the Galactic plane have revealed
several  hundred new HII regions, more than 50 new candidate supernova remnants
(SNR)  \citep{giveon05,helfand06}, and 1500 new candidate infrared stellar
clusters \citep[e.g.][]{bica03,mercer05,froebrich07}, which are often found in
the direction of HII regions. Only a few of these infrared candidate clusters
have been  confirmed with spectro-photometric studies;  the analysis is often
restricted only to  stellar clusters and does not include the cluster
environment.  A combined study of stellar clusters and their associated
molecular clouds is  a powerful tool to understand star formation. Clusters 
appear to form in large complexes \citep[e.g.][]{smith09}. The temporal and
spatial distribution of clusters  varies from cloud to cloud
\citep[e.g.][]{homeier05,kumar04,clark09b}, indicating that  external and
internal triggers are both   at work. Supernova explosions may  trigger 
subsequent  episodes of star formation in the same cloud.  The presence of SNRs 
indicates that a cloud has already undergone massive star formation, and  the
study of stellar clusters associated with  SNRs   can shed light on the initial
masses of the supernova progenitors, and therefore on the fate of massive stars.

By locating new HII regions, and young stellar clusters, we also obtain information on large scale
Galactic structures.   So far, we have only located  clusters that reside in the near side of the
Galaxy, with a few exceptions, e.g. W49 \citep{homeier05}.  Many issues on Galactic structures are
still open, e.g.\  the exact number of spiral arms,  the lack of star formation in the central 3
kpc, and the possible existence of a ring of massive star formation surrounding the central bar. 
Stellar clusters selected from infrared observations are promising tracers for Galactic studies
because they sample a larger  portion of  the Galactic plane than those from optical surveys
\citep{messineo09,davies07,figer06,clark09}.


The candidate cluster number 9 in the list by  \citet{mercer05} (hereafter, GLIMPSE9) is an
ideal target to study the issues above mentioned, because it is located in projection on  a
GMC that hosts several HII regions, supernova remnants, and  candidate  clusters. So far,
we know only of one other Galactic cluster associated with  a SNR \citep{messineo08}.
This complex is on the Galactic plane, at an heliocentric distance of  $\sim 4.9$ kpc 
\citep{albert06,leahy08}, and longitude  l=$\sim 23$\degr, and represents  an episode of
massive  star formation in the direction of the inner Galaxy.  Here we present a
spectro-photometric study of the GLIMPSE9 cluster.  

In Sect.\ \ref{observation}, we present the available photometric and spectroscopic data,
and the process of data reduction. The spectral analysis and color-magnitude diagram study
are given in Sect.\ \ref{analysis}. Information about the parent GMC and other candidate
stellar clusters are presented in  Sect.\ \ref{surrounding}. Finally, in Sect.\
\ref{summary} we summarize the results of our investigation.

\section{Observation and data reduction}
\label{observation}

\subsection{NICMOS/HST observations}  
\begin{figure}[!]
\begin{centering}
\resizebox{0.9\hsize}{!}{\includegraphics[angle=0]{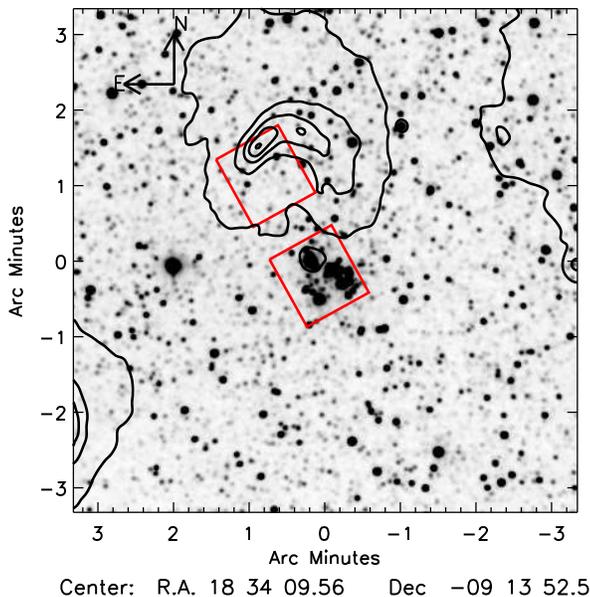}}
\end{centering}
\caption{\label{hstfields} The greyscale above shows a 2MASS \Ks-band image,
while contours show  24 \um\ emission from MIPSGAL. 
Contour levels  are 100, 200, 300, 400, and 500 MJy/sr.  The two boxes 
indicate the location and size of the two NIC3 fields. Extended emission
at a 5 $\sigma$\ level is seen towards the field.}
\end{figure}

Images were taken  with NICMOS on board  the HST on July
9, 2008 as part of the GO program 11545 (P.I. Ben Davies).  Two fields were observed; one
centered on the GLIMPSE9 cluster (RA= 18\fh34\fm09.89\fsa, DEC=-09\fdg14\farcm04.8\farcsa) and another
located at (RA=18\fh 34\fm 12.76\fsa, DEC=-09\fdg12\farcm 45.4\farcsa). The latter, which we
call the control field,  was imaged in order to  study the  background and foreground  stellar
population in the direction of the cluster.  The location of the two fields is shown
in Fig. \ref{hstfields}.  The NIC3 field of view (51.5\arcsec $\times$ 51.5\arcsec) covers the
cluster to  a radius  roughly equal to two half light radii ($\sim$30\arcsec, as measured on
2MASS images).

We used the  NIC3 camera with the F160W and F222M broadband filters, and with the F187N
and F190N narrowband filters. The fields were dithered by 5.07\arcsec\  in a spiral 
pattern (6 positions). The STEP2 sequence of the MULTIACCUM readout mode with 13 reads 
was used  for  exposures with the F160W filter, giving an integration time of 19.94 s
per exposure; the STEP8 sequence with 12 reads was used  for  exposures with the F222M
filter (55.94 s), and  the STEP8 with 10 reads for exposures with the F187N and F190N
filters (39.95 s).

\begin{figure}[!]
\begin{centering}
\resizebox{0.7\hsize}{!}{\includegraphics[angle=0]{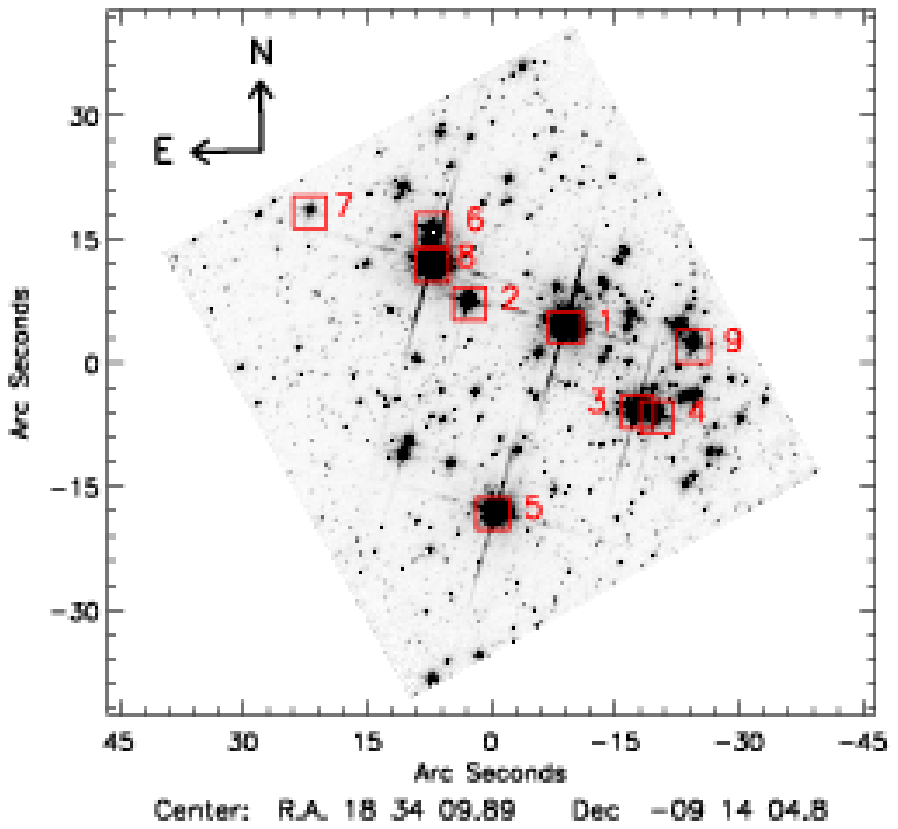}}
\end{centering}
\begin{centering}
\resizebox{0.7\hsize}{!}{\includegraphics[angle=0]{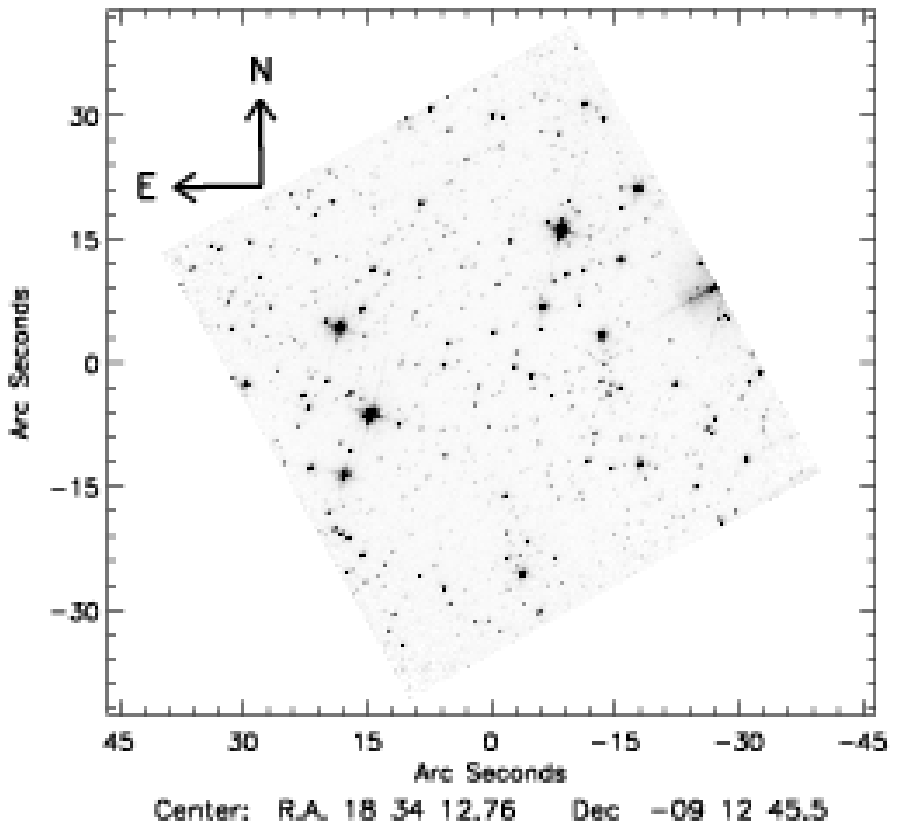}}
\end{centering}
\caption{\label{hstmosaic} {{\bf Top:} F160W mosaic of the GLIMPSE9 cluster.
Squares  and  identification numbers (from  Table \ref{table.targets}) 
show the location of spectroscopically observed stars. Stars \#10 and \#11
fall outside the field of view. 
{\bf Bottom:} F160W mosaic of the control field.} }
\end{figure} 

\subsection{NICMOS/HST data reduction}   
The images were bias subtracted, dark corrected, and flat-fielded  by the
standard NICMOS pipeline CALNICA (see the NICMOS Data Handbook
v7.0\footnote{$http://www.stsci.edu/hst/nicmos/$}). The six dithered exposures
of each observation were re-sampled into  a final mosaic with a pixel scale of
0.066\arcsec.  The cluster mosaics are shown in  Fig.\  \ref{hstmosaic}.

A major photometric uncertainty is inherent for images with the NIC3 camera, and
is due to the combination of an under-sampled point spread function (PSF) with
the lower intrapixel sensitivity of the camera (up to 10-20\%, see the NICMOS
Data Handbook v7.0). The dithering observing strategy reduces this uncertainty.
A photometric analysis of the mosaics was carried out with DAOPHOT
\cite{stetson87} within the Image Reduction and Analysis Facility (IRAF). We
used the images of the control field, which is less crowded, to build a
point-spread-function (PSF); seven  isolated and bright stars were selected in
the F160W and F222M images.   Due to the small number of isolated stars no
attempt was made to model a spatially varying PSF.   Aperture photometry with a
radius of 1.1\arcsec\ was performed  on  bright and isolated stars, and the
average difference between these magnitudes and the PSF-fitting magnitudes from
DAOPHOT was measured. This aperture correction was applied to the whole
catalogue. We obtained calibrated magnitudes in the Vega system with the
transformation equations between counts/s  to Vega magnitudes as given in the
NICMOS Data Handbook.   Astrometry and photometric calibrations into the 2MASS
system were obtained  using a set of 13  point sources with good quality $H$ and
$K_S$ measurements in 2MASS \citep{skrutskie06}. The reference stars span a 
$H-K_S$ range from 0.2  to 1.4 mag, and a $K_S$ range from 6.3 to 14.0 mag.

\begin{figure}[!]
\begin{centering}
\resizebox{0.9\hsize}{!}{\includegraphics[angle=0]{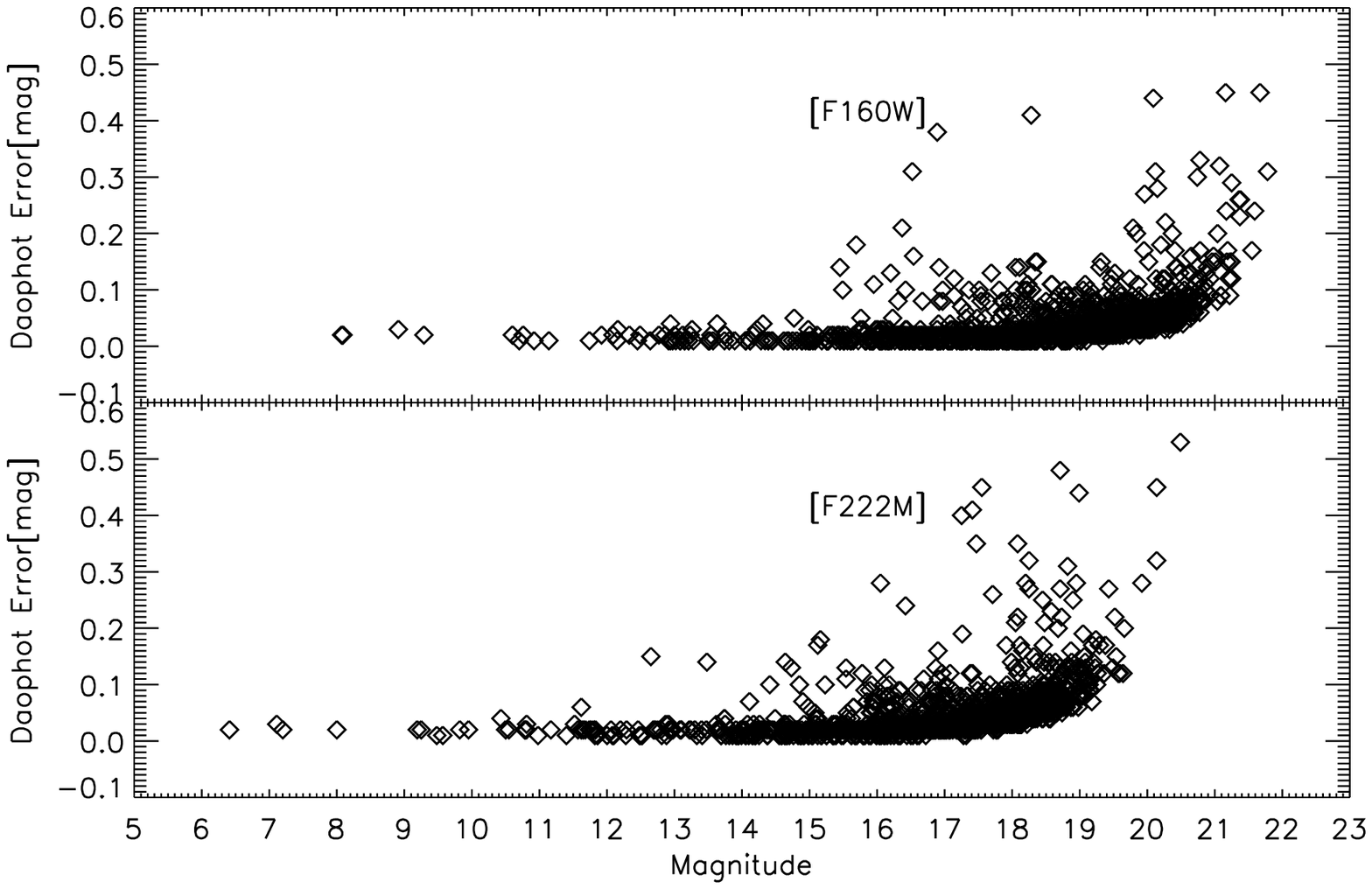}}
\end{centering}
\caption{\label{hstcalib} {Magnitude errors from DAOPHOT for the cluster field
([F160W] in the upper panel, and [F222M] in the lower panel).} }
\end{figure}

The transformation equations are:
{\small
\begin{eqnarray*}
K_s-[F222M]=~~~~~~~~~~~~~~~~~~~~\\
(-0.01\pm0.10)+(-0.03\pm0.08)\times([F160W]-[F222M]),
\end{eqnarray*}

\begin{eqnarray*}
H-[F160W]=~~~~~~~~~~~~~~~~~~~~ \\
(-0.08\pm0.09)+(-0.22\pm0.07)\times([F160W]-[F222M]), 
\end{eqnarray*}
}

with a standard deviation of 0.11, and 0.10 mag, respectively. $[F222M]$  and  $[F160W]$ 
are magnitudes  in the VEGA photometric system (following the NICMOS manual), and 
$K_S$  and $H$ are magnitudes in the 2\,MASS system. A plot of the photometric 
errors given by DAOPHOT is shown  in Fig.\ \ref{hstcalib}.

\begin{figure}[!]
\begin{centering}
\resizebox{0.9\hsize}{!}{\includegraphics[angle=0]{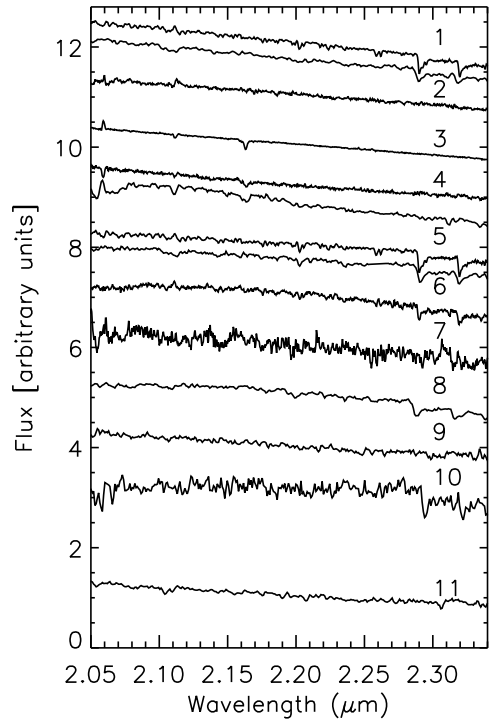}}
\end{centering}
\caption{\label{spectra} {\bf Left panel} K-band spectra taken with NIRSPEC and
IRMOS.  The number labels refer to  Table \ref{table.targets}. The spectra have 
been corrected for interstellar extinction using \Aks=1.5 mag and the extinction 
law by \citet{messineo05}. For  tree stars (\#1, \#4, and \#5) with both KECK and 
IRMOS observations both spectra are shown (the Keck spectrum is the top one).}
\end{figure}

We also extracted point sources from the F187N and F190N  images.
The difference of [F187N]  and [F190N]  magnitudes show a scatter of 0.3 mag.
Within this uncertainty, no emission lines were detected.

{
\begin{deluxetable*}{rlllllrrr}
\tablewidth{0pt}
\tablecaption{\label{table.targets} List of stars spectroscopically observed.}
\tablehead{
\colhead{ID}& 
\colhead{RA}& 
\colhead{DEC} & 
\colhead{frame} & 
\colhead{Spec. type} &
\colhead{EW(CO)} &
\colhead{H} &
\colhead{\Ks} &
\colhead{\Aks} 
}
\startdata
 1 &18 34 09.266 & $-$09 14 00.74&{\small NIRSPEC}(118r)/{\small IRMOS}(ir3)& M1-0I  &   -43-39 & 07.80 &07.17&$0.6\pm0.1$\\
 2 &18 34 10.064 & $-$09 13 57.74&{\small NIRSPEC}(118l)                    & $<$G   &  \nodata & 10.30 &09.20&$1.6\pm0.1$\\
 3 &18 34 08.692 & $-$09 14 11.07&{\small NIRSPEC}(120l)                    &   OBI  &  \nodata & 08.93 &07.96&$1.3\pm0.1$\\
 4 &18 34 08.537 & $-$09 14 11.83&{\small NIRSPEC}(120r)/{\small IRMOS}(ir4)&	OB   &  \nodata & 10.21 &09.14&$1.5\pm0.1$\\
 5 &18 34 09.857 & $-$09 14 23.28&{\small NIRSPEC}(122)/{\small IRMOS}(ir2) & M1-2I  &   -43-49 & 08.43 &07.05&$1.7\pm0.1$\\
 6 &18 34 10.353 & $-$09 13 48.99&{\small NIRSPEC}(124r)                    & M5III  &	 -32    & 10.41 &09.44&$1.0\pm0.1$\\
 7 &18 34 11.348 & $-$09 13 46.47&{\small NIRSPEC}(124l)                    & $<$G   &  \nodata & 12.05 &10.93&$1.6\pm0.1$\\
 8 &18 34 10.352 & $-$09 13 52.95&{\small IRMOS}(ir1)                       & M2.5I  &	 -49    & 07.58 &06.30&$1.5\pm0.1$\\
 9 &18 34 08.228 & $-$09 14 03.25&{\small IRMOS}(ir5)                       & $<$G   &  \nodata & 10.77 &09.78&$1.4\pm0.1$\\
10\tablenotemark{a} & 18 34 06.662 & $-$09 14 47.40&{\small IRMOS}(ir6)     & $>$K0  &  \nodata & 12.26 &10.26&$2.7\pm0.1$\\
11\tablenotemark{a} & 18 34 05.345 & $-$09 14 24.01&{\small IRMOS}(ir7)     & $<$G   &  \nodata & 10.05 & 8.96&$1.6\pm0.1$\\ 
\enddata		 
\tablenotetext{a}{ 2MASS magnitudes are listed for stars N. 10 and 11, because they are outside the area covered by the NIC3 mosaic.}
\tablecomments{For each star, number designations and coordinates
(J2000) are followed by the instrument name (plus frame name), 
spectral classification, EW(CO), H and \Ks\ magnitudes obtained 
from the HST images, and the estimated interstellar extinction (in magnitude). }
\end{deluxetable*}
}

\subsection{Spectroscopic data}    We obtained spectroscopic observations  with the Infra-Red
Multi-Object Spectrograph (IRMOS) at the Kitt Peak Mayall 4m telescope on September 25th,
2007 \citep{mackenty03}. We used the K1000 grating in combination with the K filter to cover
the wavelength  region from 1.95 \um\ to 2.4 \um\  with a resolution of R=$\sim1000$. A
number of 28 exposures of 1 minute each were taken in two nodded positions. In order to
remove variable background signals each exposure was followed by a dark observation  of equal
integration time. Neon lamp and continuum lamp observations were taken soon after the target
observations. Dark-subtracted science frames were combined and flat-fielded. A two
dimensional de-warping procedure was then used to straighten the stellar traces before
extraction. Wavelength calibration was obtained by using both  neon lines and OH lines
\citep{oliva92}. A total of seven spectra were extracted. Each target spectrum was divided by
the spectrum of an A2V star in order to correct for atmospheric absorption  and instrumental
response. The \brgamma\ of the telluric spectrum was eliminated  with  linear
interpolation.  The intrinsic shape of the telluric spectrum was  removed
by dividing it by a black body function of 9120 K \citep{blum00}.

Additional spectroscopic observations were carried out with NIRSPEC at the KeckII telescope  under
program U050NS (P.I. M. Rich), on July 11th, 2008. We used the K filter and a 42\arcsec $\times$
0.570\arcsec\ slit.  A  wavelength coverage from 2.02 \um\ to 2.45 \um\ and  a
resolution R=1700 were obtained. For each star, two exposures were taken of 10s each, in two nodded
positions along the slit. We used a continuum lamp observation as a flat field, and Ar, Ne and Kr lamps
observations for the wavelength calibration. Pairs of nodded positions were subtracted and
flat--fielded. Atmospheric absorption and instrumental response were removed by dividing each extracted 
target by the spectrum of a B0.5V telluric standard (HD1762488), and multiplying for a blackbody
spectrum of 32,060 K \citep{blum00}. A total of seven spectra were extracted from the NIRSPEC
observations.  Three of the stars observed with IRMOS were  also observed with NIRSPEC.

A chart of the observed stars is shown in Fig.\ \ref{hstmosaic}. The seven spectra with
NIRSPEC and the additional four spectra with IRMOS are shown in Fig.\  \ref{spectra}. In
Table \ref{table.targets} we list the coordinates of the stars  spectroscopically observed.

\section{Analysis}
\label{analysis}

\subsection{Spectral types}  

Spectral classification was performed by comparing the spectra with spectral atlases
\citep[e.g.][]{hanson96,hanson05,ivanov04, alvarez00,  blum96, kleinmann86, wallace96}.
$K$-band observations enabled us to classify both late- and early-type stars. Typically,
spectra of late-type stars show CO  band-head at 2.29 \um\ and   atomic lines from Mg I, Ca I,
and Na I;  early-type stars can be identified by detecting  hydrogen (H) lines, helium (He)
lines, and other atomic lines, e.g.\  CIV triplet at 2.069, 2.078, and 2.083 $\mu$m, and 
broad emission  at 2.116 $\mu$m, which is  due to NIII, CIII and HeI emission.

\begin{figure}[!]
\begin{centering}
\resizebox{0.9\hsize}{!}{\includegraphics[angle=0]{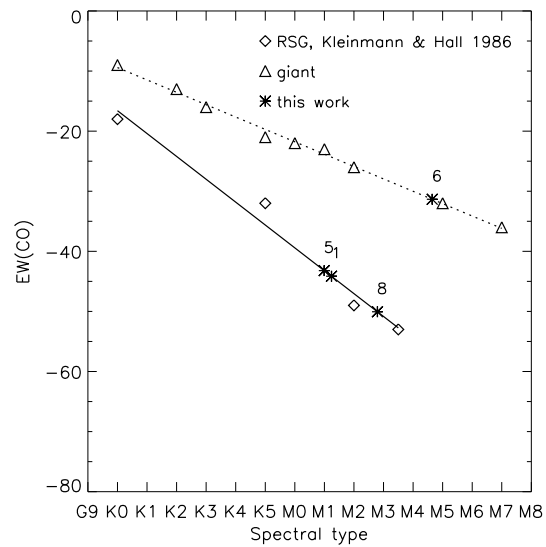}}
\end{centering}
\caption{\label{ew} Measurements of  EW(CO) versus spectral type for the
late-type stars identified in GLIMPSE9 (asterisks).
For comparison,  known giant stars (triangles) and  RSG stars (diamonds)
are also shown \citep{kleinmann86}. The lines indicate a linear fit to the
EW(CO) versus spectral type of giants (dotted) and RSG stars (continuous).
There is a small  difference between the effective temperature of a K5 and 
an M0 star ($\sim$150K).
Further subclassification using the EW(CO) are not possible.
}
\end{figure}

Stars \#1, \#5, \#6, \#8, and \#10 show CO bands in absorption, which indicate low effective
temperatures. Since  the absorption strength  of CO band-heads increases with decreasing effective
temperature $T_{\rm eff}$, but with increasing luminosity L, giant stars and supergiant stars  follow 
different equivalent width EW(CO) versus temperature relations  \citep[e.g.\ ][]{davies07}. For each
star,  we measured the EW(CO) band-head feature   between 2.285 \um\ and 2.315 \um, with an adjacent
continuum measurement made at 2.28--2.29 \um. Then, we compared these measurements to those of template
stars \citep{kleinmann86}, and determined the spectral type (see Fig. \ref{ew}). Stars \#1, \#5 and \#8
appear to be red supergiant stars (RSGs) with spectral types between M0 and M2.  Star \#6 is most likely a giant M5 because of
its small EW(CO) and  of its lower luminosity (\Ks = 9.4 mag). Star \#10 also shows CO band-head in
absorption, but because of poor  signal-to-noise ($\sim10$) no spectral typing is attempted.

Stars \#3 and \#4 show a \brgamma\ line and a weak HeI line at 2.11 \um\ in
absorption. A HeI line at  2.05 \um\ is in emission in spectrum of \#3, while in
absorption in the spectrum of \#4. From a comparison of the spectra with those given 
in \citet{bibby08} and \citet{hanson96},  star \#3  appears to be a B1-B3 supergiant.
Star \#4 is probably earlier than \#3 (O9-B0) because it is fainter in the \Ks-band and 
the \brgamma\ absorption is weaker.

The spectra of stars \#2, \#7, \#9, and \#11  do not show any lines, however, the
absence of CO band-head suggests a spectral type earlier than G type. 

\subsection{Color-magnitude diagrams} 
\begin{figure}[!]
\begin{centering}
\resizebox{1\hsize}{!}{\includegraphics[angle=0]{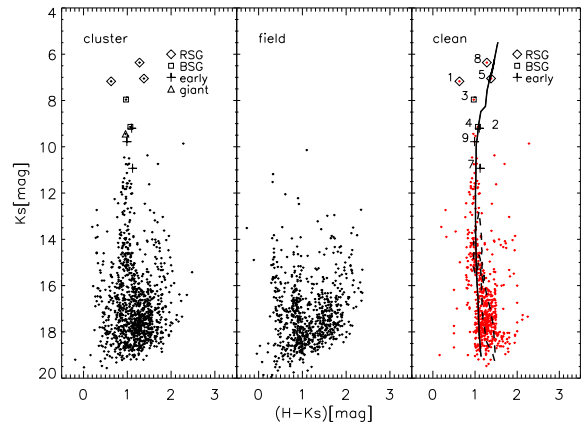}}
\end{centering}

\caption{\label{hstcmd} {\bf Left panel:} HST/NICMOS $H-$\Ks\ versus \Ks\ color  magnitude
diagram of point sources extracted from the cluster images.  {\bf Middle panel:} Similarly,
a color magnitude diagram for the control field (of equal area) is shown.  {\bf Right
panel:} A color-magnitude diagram of the cluster field after statistical decontamination.
The vertical line indicates an isochrone of 15.0 Myr and solar metallicity \citep{lejeune01}, which was
shifted to a reddening of \Aks=1.5 mag and  a distance of 4.2 kpc \citep{leahy08}. Dotted
lines show isochrones corresponding to a population of 0.5 Myr, and 3 Myr with solar
metallicity \citep{siess00}. The diamond symbols indicate the location of  RSGs.  Squares
indicate  BSGs, and a plus symbol a candidate early type star. A triangle shows a giant
star, which is likely unrelated to the cluster. Stars are numbered as in Table \ref{table.targets}. } 
\end{figure}

Color-magnitude diagrams (CMD) of  HST/NICMOS point sources are presented in Fig.\
\ref{hstcmd}.  The CMD of the cluster field  presents a clear sequence of stars with $H$-\Ks$=\sim$1
mag and  \Ks$<15$ mag, while only a few stars populate the same region  of the CMD for point
sources extracted from the control field. The control and cluster fields were observed in the
same  way,   and therefore have identical area. 

Magnitudes in the 2MASS system are preferred for the CMD in order to have a direct comparison with the
CMDs in \citet{messineo09}, \citet{figer06}, \citet{davies07}.

To isolate the cluster sequence, we performed a statistical decontamination using stellar
counts per 0.5 mag bin of ($H-$\Ks)  color and 1.0 mag bin of \Ks\ magnitude in both 
cluster and field regions;
we randomly subtracted from each cluster bin a number of stars equal to that of the
corresponding field bin.  The resulting diagram is shown in the right panel of Fig.\
\ref{hstcmd}.

There is a kink in the ''clean'' CMD at \Ks\ $\propto 15-16$ mag. 
This could be due to a poor field subtraction, or to a pre-main sequence. 
If we presume the faint stars to be a pre-main sequence, then their  ages would range  between 0.5 
and 3 Myr (Fig.\ \ref{hstcmd}). NGC7419, which contains 5 RSGs and has a age of about 10 Myr,
also shows a younger sequence \citep{subramaniam06}. Further observations are needed to study 
the nature of these faint stars.

From the CMD, we estimate an interstellar  extinction of \Aks$=1.6\pm0.2$ mag by measuring the
median $H$-\Ks\ color of cluster   with $K_S < 15$ mag and $0.6 <$ $H-$\Ks $< 1.4$ mag, and by adopting the
extinction law by \citet{messineo05}.  This measurement is independent of age because in the $H$-\Ks\
versus \Ks\ diagram,  the isochrones are almost vertical lines.

We spectroscopically detected several massive stars: three RSGs, with spectral type  from
M0 to M2, and two blue supergiant stars (BSGs) (O9-B2). Since RSG stars span a broad range of magnitudes, they
cannot be used  as a distance indicator.  Therefore, to determine the  cluster distance and
other dependent parameters (e.g. age and mass),  one must rely on stars other than RSGs. 
For the BSGs \#3 and \#4 we measured an interstellar extinction  \Aks$=1.3$ mag and 1.5
mag, respectively  \citep[using intrinsic magnitudes and extinction law
from~][]{bibby08,messineo05}.  These extinction values are consistent with that of the
cluster sequence, and suggest membership.  For star \#3 we  obtained a spectrophotometric
distance of $4.2\pm0.4$ kpc, and  for star \#4 of $4.7\pm0.4$ kpc. These values are
consistent with  a distance of $4.2\pm0.3$ kpc, which was inferred for the GMC and 
the SNR W41 by \citet{leahy08}.   In the following we will adopt a distance of 4.2 kpc.

The three brightest stars ($6.3 <$ \Ks $<7.2$ mag) have values of EW(CO) typical of RSGs. By
comparing their  observed  colors with  intrinsic colors of RSGs \citep{koorneef83}, and using
the  extinction law given in \citet{messineo05}, we estimated the following   values of
interstellar extinction:  \Aks$=0.6\pm0.1$ mag for star \#1 (\Ks=7.17 mag), \Aks$=1.5\pm0.1$ mag 
for star \#8 (\Ks=6.3mag),  and  \Aks$=1.7\pm0.05$ mag for star \#5 (\Ks=7.05 mag).  Stars \#5 and 
\#8 have extinction values consistent with those of  the cluster sequence, and are likely members,
while  the bluer color of  star \#1 suggests a foreground star. 

Using the relation between effective temperature and  bolometric correction for RSGs given by
\citet{levesque05}, and a distance of 4.2 kpc, we derived a bolometric luminosity
Mbol=$-5.7\pm0.1$ mag for star \#8 and Mbol=$-5.1\pm0.1$ mag for star \#5. From non-rotating
evolutionary tracks with Solar abundance by \citet{meynet94}, we inferred   masses of $10.5\pm1.5$ \Msun, 
which corresponds to an age of $22.5\pm4.5$ Myr. Similar range is obtained when using
the newer non-rotating models by \citet{meynet03}. By increasing the distance by 40\% (6 kpc)
the minimum age would decreases by  $\sim$ 30\% (15 Myr).  

Star \#8 shows  water absorption  at the blue edge of the $K$-band (1.9-2.1 \um).  Its  EW(CO) argues   for a late
giant type (M7III), or  an M2 RSG. The water absorption indicates the presence of a circumstellar
envelope. Star \#8 has  a  \Ks$-[8]=1.5$ mag, redder than the 0.42 and 1.14 mag of star \#1 and star
\#5 (magnitudes at 8 \um\ are from the SPITZER/GLIMPSE survey).  Water is typically seen in  AGB
stars and RSG stars \citep{tsuji00,blum03}.  However, the location  in the direction of a young
cluster, and the rarity of bright infrared stars, suggests a RSG cluster  member. 
We estimated a stellar density of $0.17\pm0.1$  bright stars (\Ks$<7.5$ mag) per square arcminute, 
in the longitude range from 20\degr\ to 30\degr, and within 0.5\degr\ from the Galactic plane; 
there are three such bright stars at the location of GLIMPSE9, and 
one is likely to be a chance alignment by virtue of the bluer color.
Radial velocities are  needed to firmly solve the puzzle.

\subsection{Luminosity function}   
\begin{figure}[!]  
\begin{center}
\resizebox{0.55\hsize}{!}{\includegraphics[angle=0]{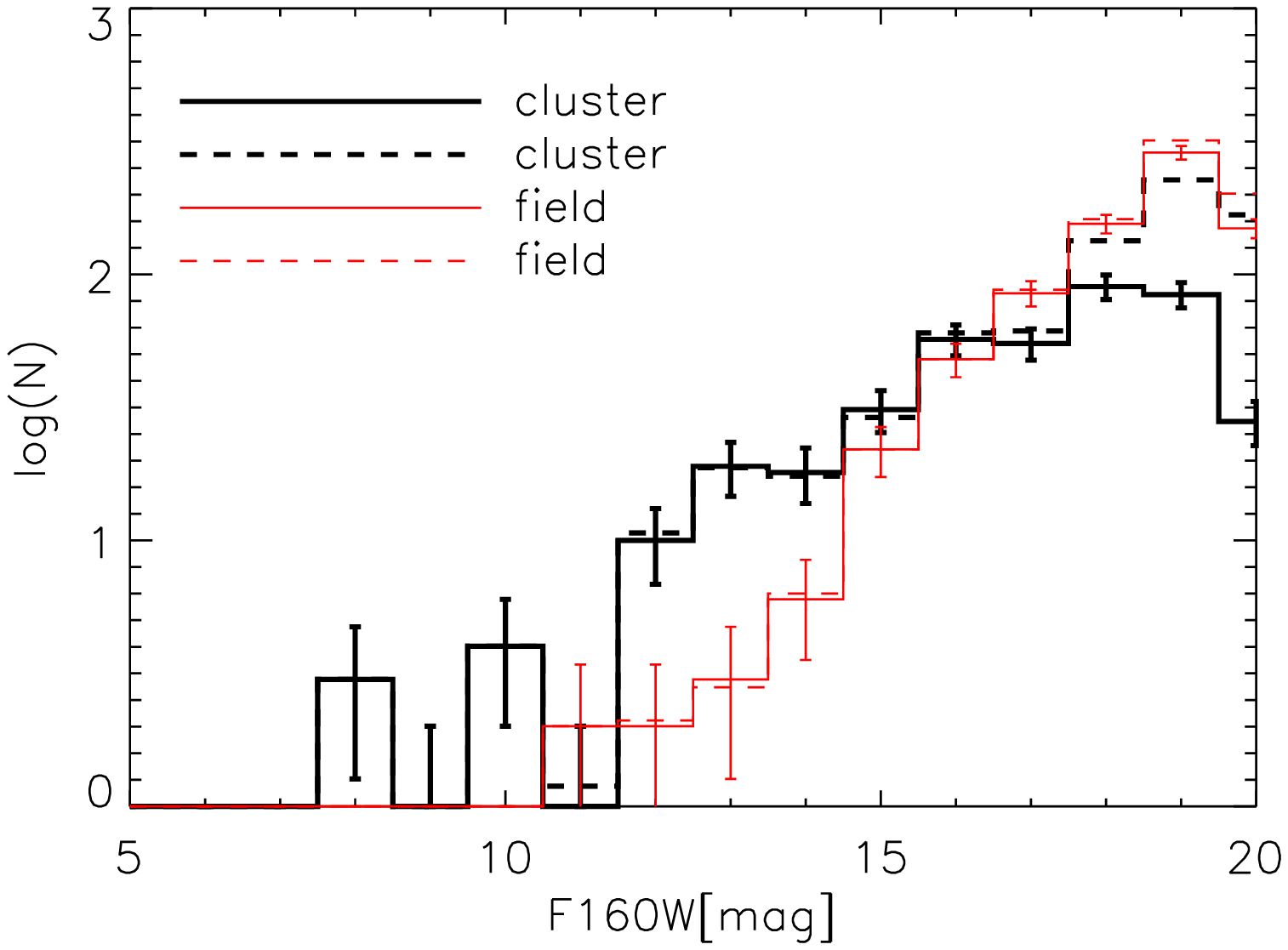}}
\end{center}  
\begin{center}
\resizebox{0.55\hsize}{!}{\includegraphics[angle=0]{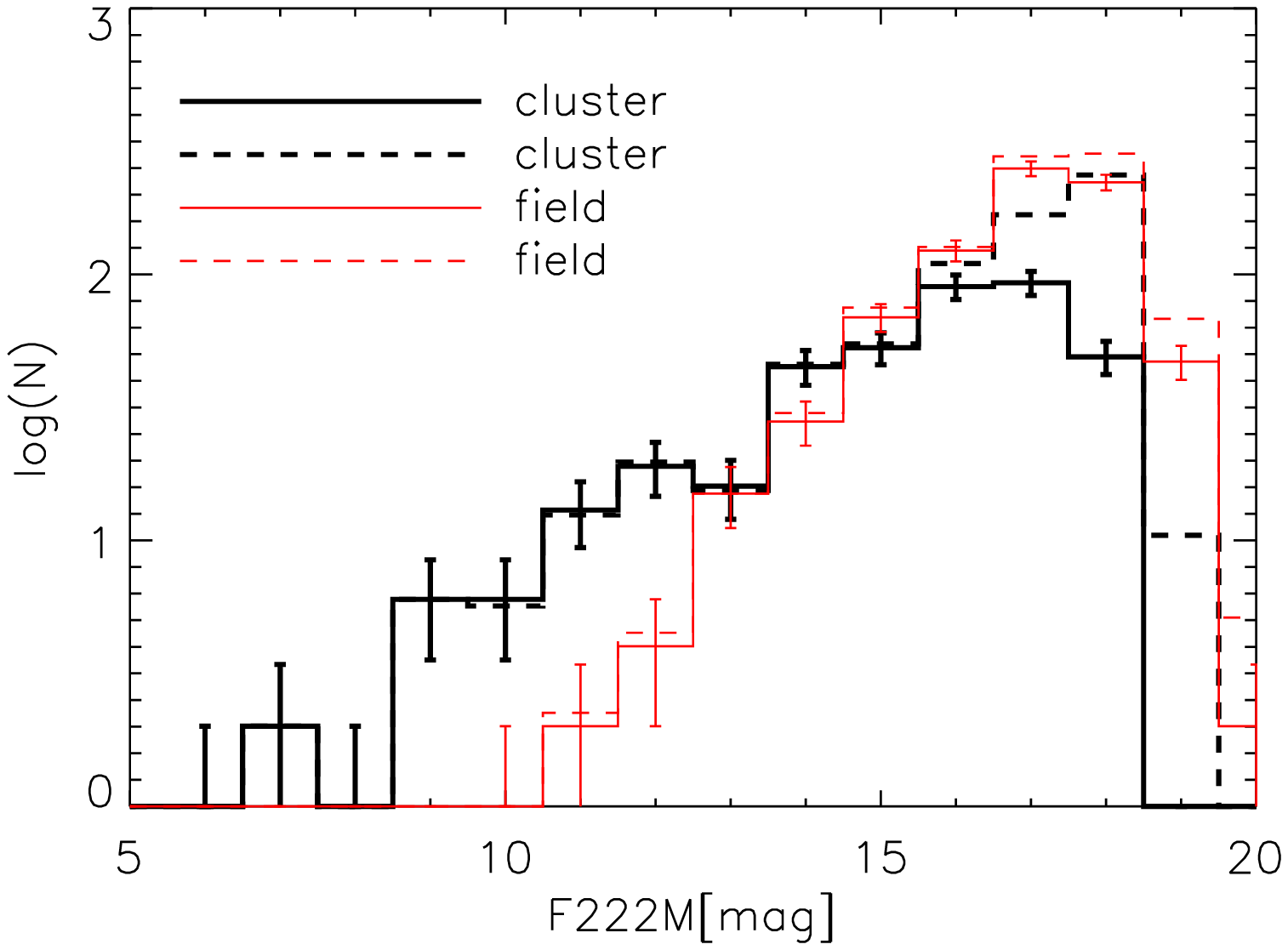}}
\end{center}  
\begin{center}
\resizebox{0.55\hsize}{!}{\includegraphics[angle=0]{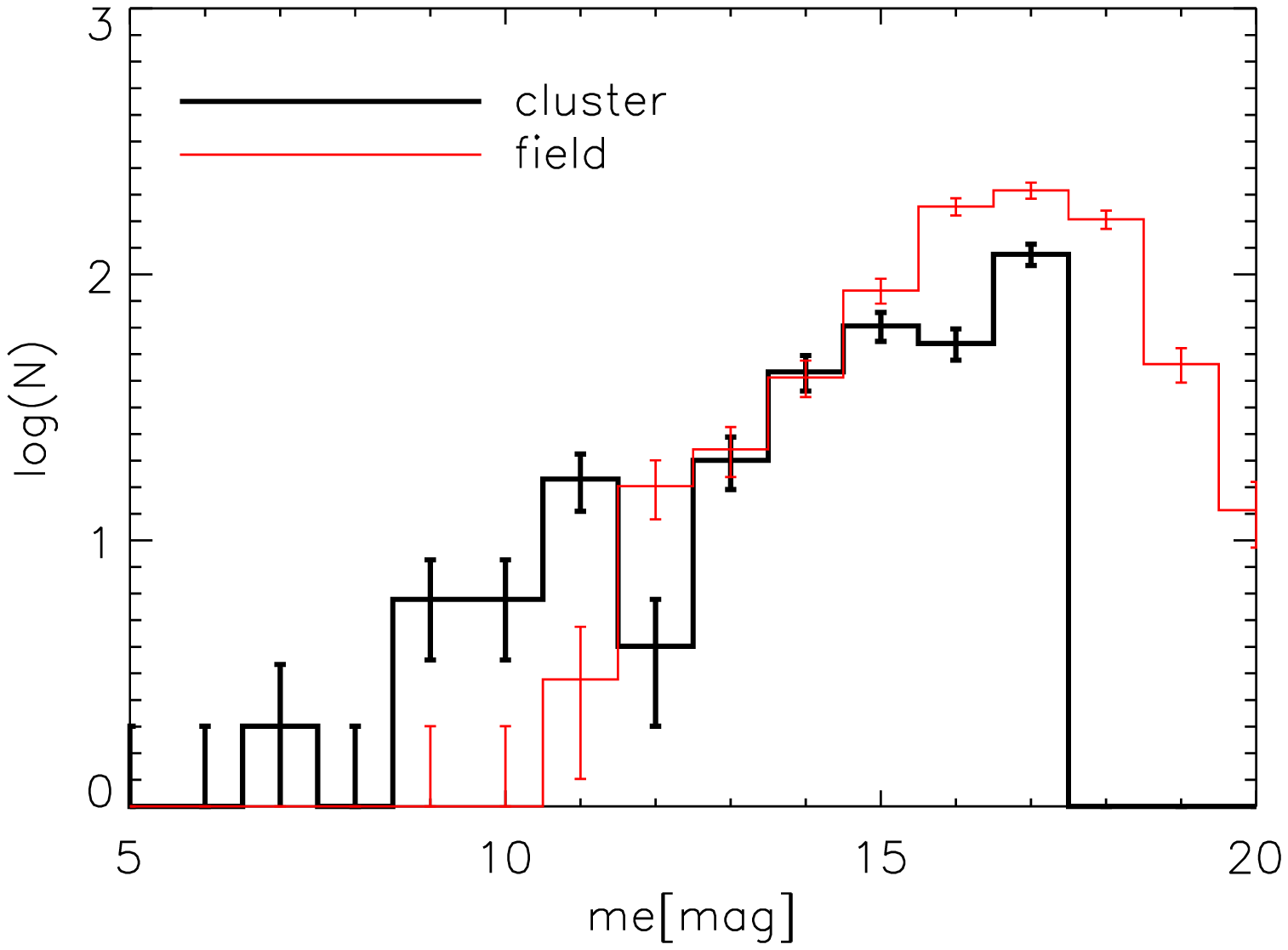}}
\end{center}  
\caption{\label{LF} Luminosity functions. {\bf Top panel:}  Observed LFs of stars detected in the
F160W mosaics. The thick continuous line shows the GLIMPSE9 cluster  LF after field
subtraction, while the thick dashed line shows the same LF after correcting for
incompleteness. Similarly, the thin  lines show the LFs of the control field without
(continuum) and with (dashed) correction for incompleteness.  {\bf Middle panel:}  
Observed LFs of stars detected in the F222M mosaics.  {\bf Lower panel:}  
LFs with the extinction free magnitude $\rm m_{e}$ of stars detected in both bands. 
Since $\rm m_{e}$ depends on color, we do not apply any extinction correction. 
Notice that Glimpse 9 has more luminous (massive) members.}  
\end{figure}

The cluster and field  LFs of stars detected with the  F160W filter are shown in the top panel 
of Fig.\  \ref{LF}, while that of stars detected with the F222M filter are in the middle panel 
of the same figure. Since the transformations to the 2MASS system depend on a color term, the VEGA 
photometric system must be used to independently analyze the individual bands.
The  cluster LF is estimated by subtracting the  field stars from the observed LF, and
dividing it by the  completeness factor (see the Appendix). The cluster LF shows an evident excess of stars with
[F222M] brighter than $\approx 15$ mag.

MIPSGAL data reveals that the control field has increased 24\um\ background
emission compared to the cluster field (see Fig.\ \ref{hstfields}),  
indicating that it may have a slightly higher extinction. However, from the 
CMDs and LFs it appears that this does not  affect the analysis for 
stars brighter than [F222M]$<15$ mag (\Ks$<\sim15$ mag).  

We also construct a LF for an extinction free magnitude defined as 

\begin{eqnarray*}
m_e = Ks - 1.5 \times (H-Ks-1.0),
\end{eqnarray*}

where the 1.5 constant is the ratio between  interstellar extinction in \Ks-band and the  reddening
in $H-$\Ks\ \citep{messineo05}, and 1.0 mag is the average $H-$\Ks\ of the  cluster sequence. By
using $m_e$, each point source is moved on the CMD along the reddening vector to an observed
$H-$\Ks=1 mag. Because in $H-$\Ks\ the intrinsic color of stars is almost independent
of spectral type (within 0.3 mag) \citep{koorneef83}, the comparison between field and cluster luminosity functions becomes 
extinction free. If the control field would have a systematically higher reddening than the cluster
field, stars detected  in the control field would have higher $H-$\Ks\ and lower  \Ks\ than the true
foreground population of the cluster. This could cause an artificial excess of bright stars in the
cluster LF of the \Ks\ band, which in turn would bias the  mass function. The LF with $m_e$  shows a similar excess 
of bright stars (lower panel of Fig.\ \ref{LF}). Thereby, the stellar over-density at the location
of the GLIMPSE9 candidate cluster is confirmed. 

A drop appears in the LF around $m_e = 12$  mag.  This is due  to an excess of stars  with 
($H-$\Ks)$< 1.8$ mag in the control field, and disappears when applying a color selection before
building the mass function.

\subsection{Mass function}  
\begin{figure}[!] 
\begin{centering}
\resizebox{0.9\hsize}{!}{\includegraphics[angle=0]{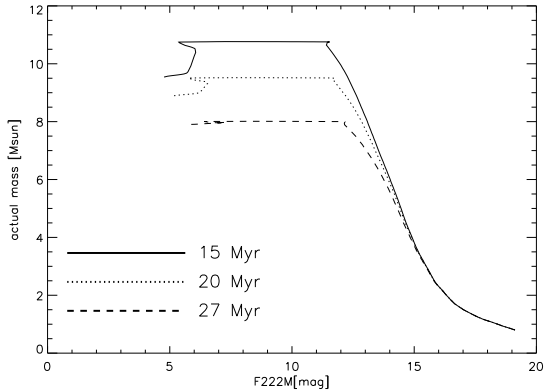}}
\end{centering} 
\caption{\label{masslum} Actual mass versus apparent [F222M], for populations
of 15, 20, and 27 Myr, with  Solar  abundance and increased mass-loss \citep{lejeune01}. 
A distance of 4.2 kpc, and \Ak=1.6 mag are used.}
\end{figure}

\begin{figure}[!] 
\begin{centering}
\resizebox{0.5\hsize}{!}{\includegraphics[angle=0]{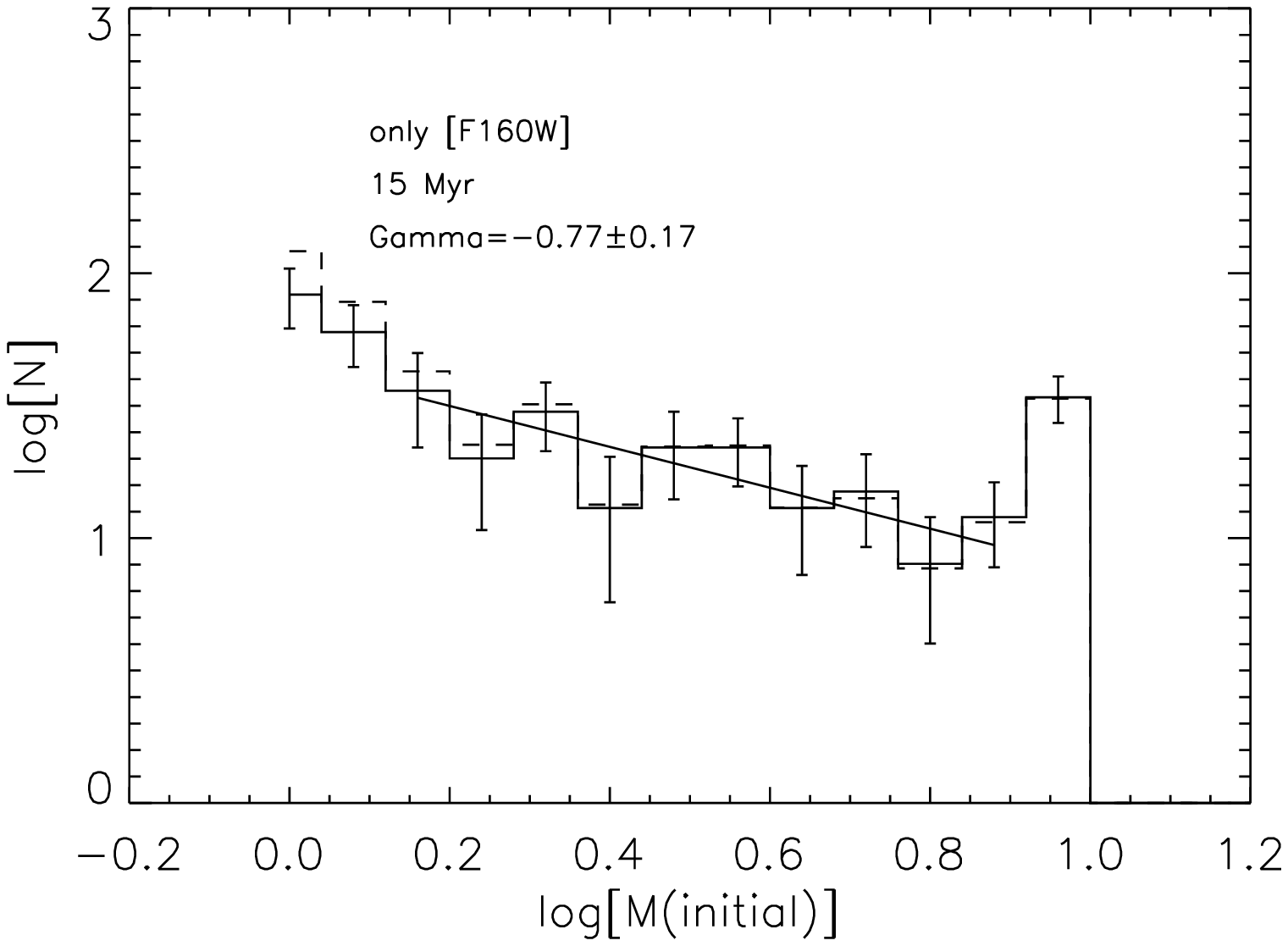}}
\resizebox{0.5\hsize}{!}{\includegraphics[angle=0]{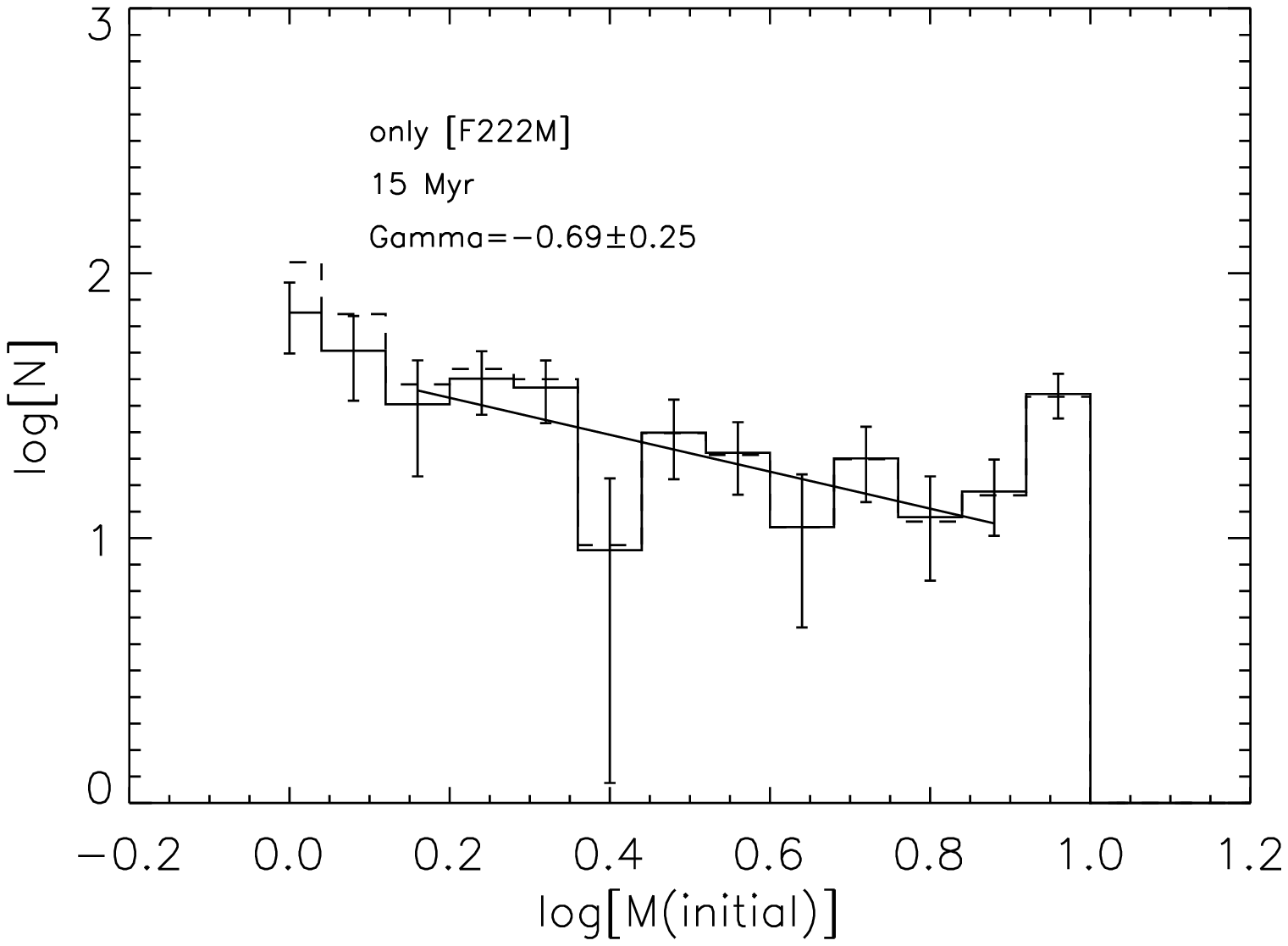}}
\resizebox{0.5\hsize}{!}{\includegraphics[angle=0]{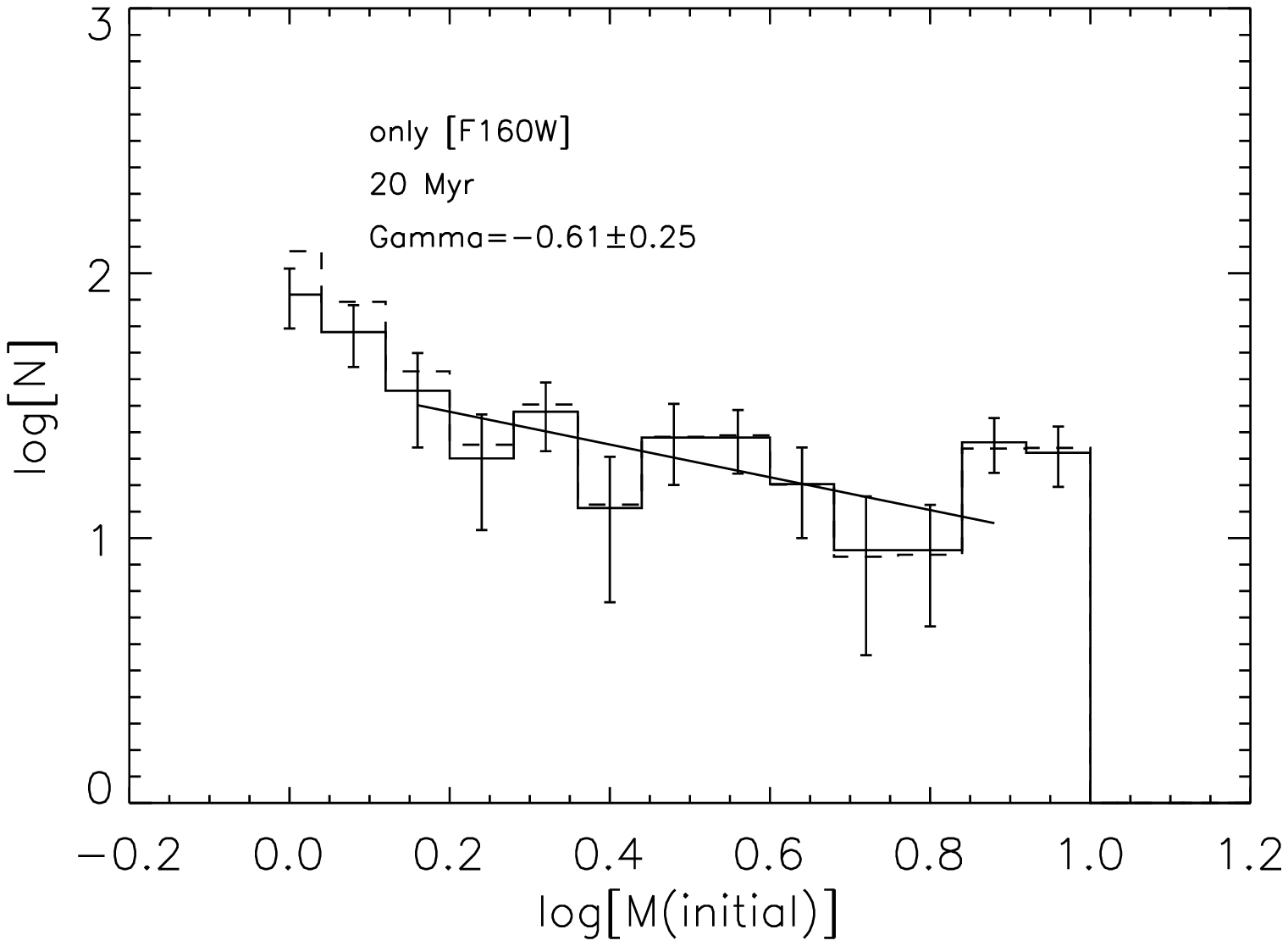}}
\resizebox{0.5\hsize}{!}{\includegraphics[angle=0]{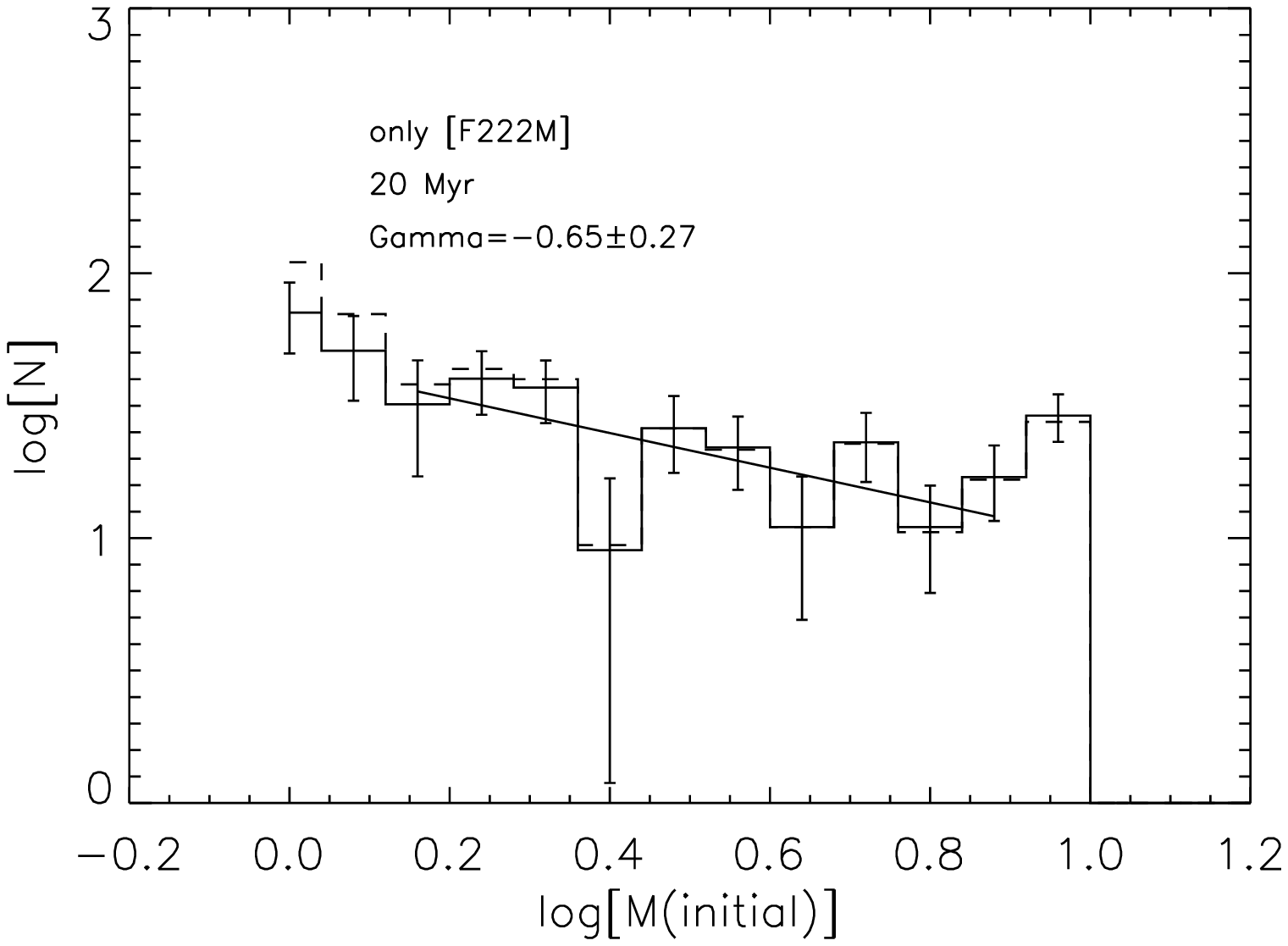}}
\resizebox{0.5\hsize}{!}{\includegraphics[angle=0]{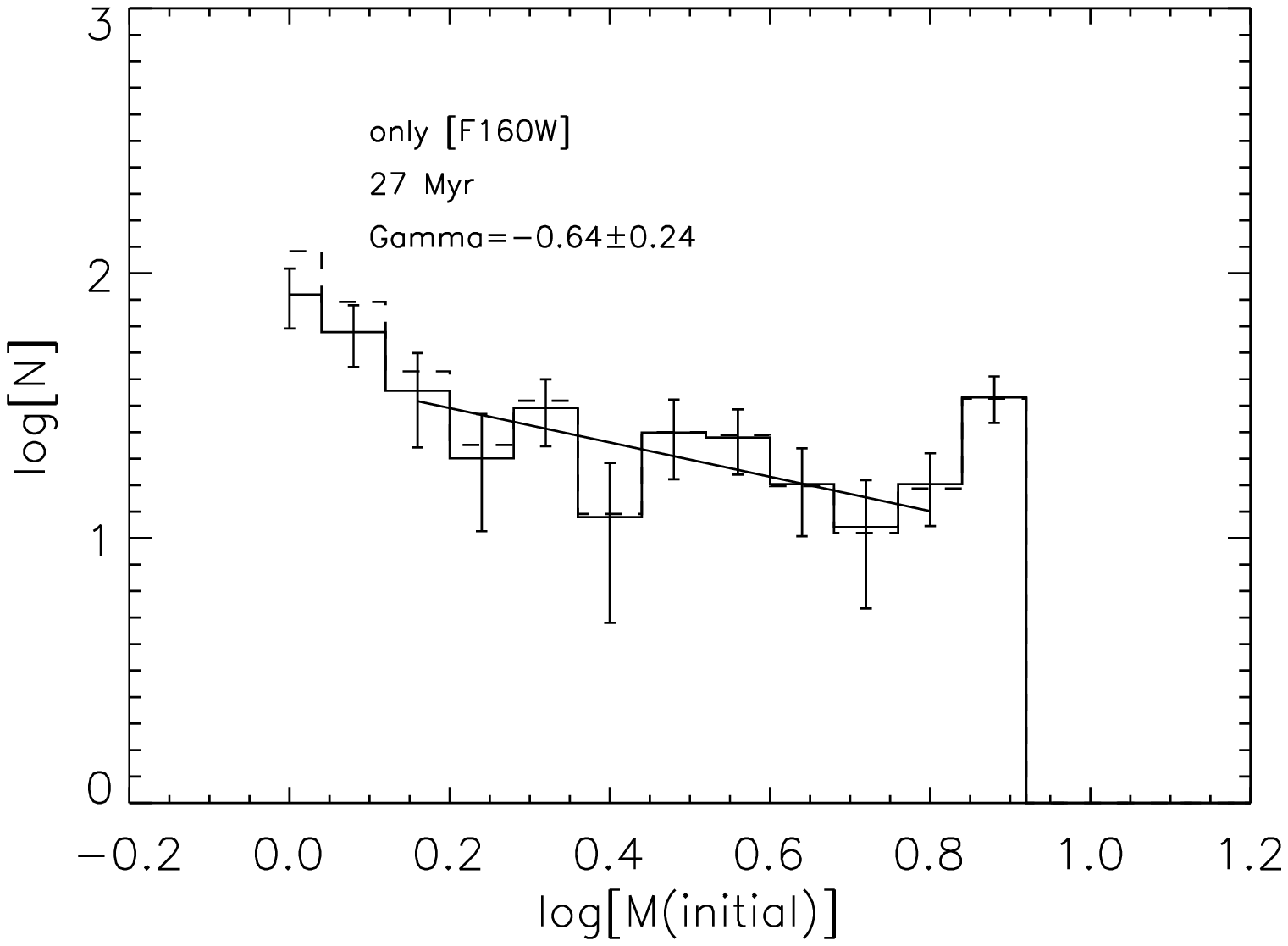}}
\resizebox{0.5\hsize}{!}{\includegraphics[angle=0]{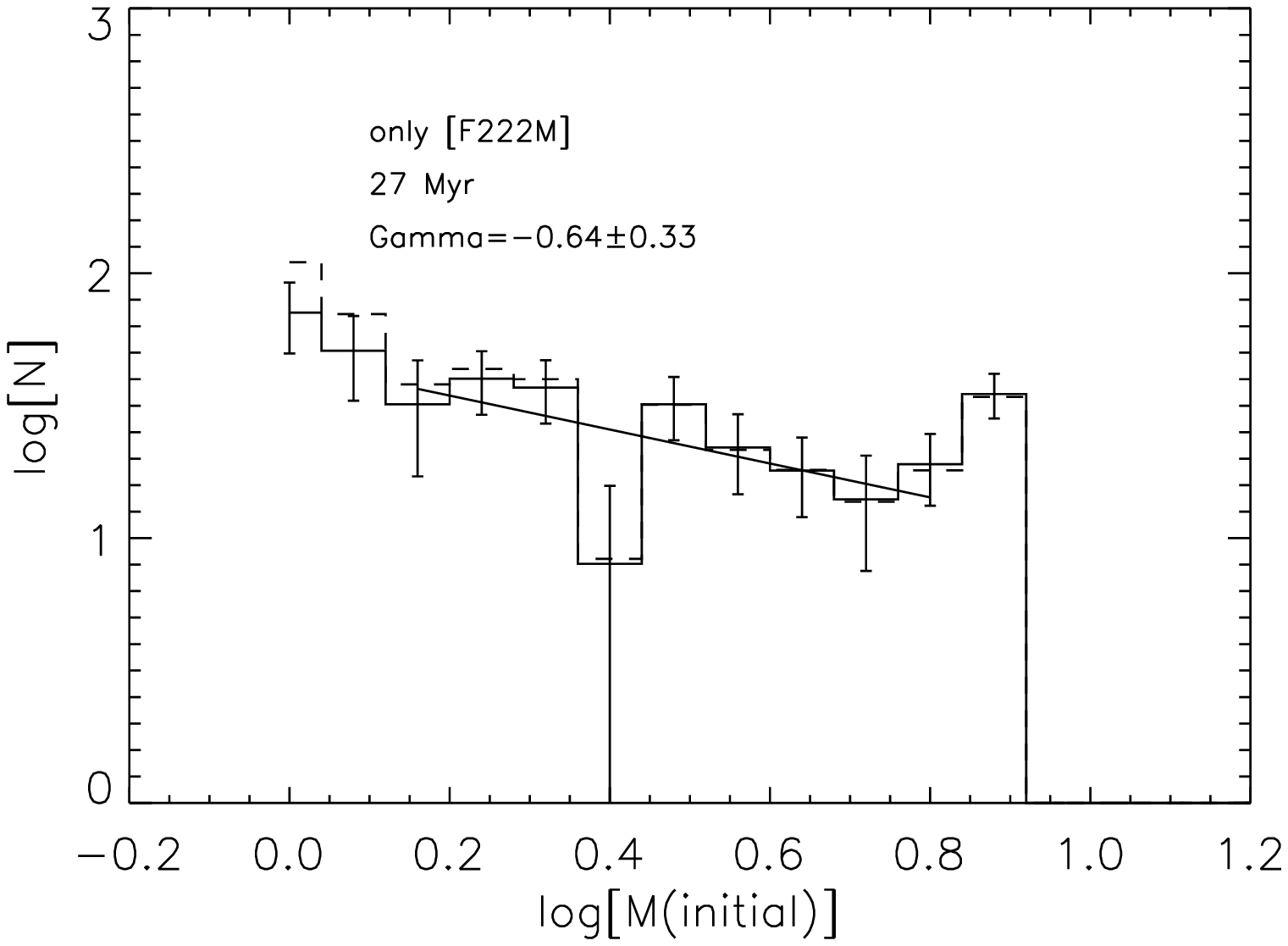}}
\end{centering} 
\caption{\label{gl09spectra} Background-subtracted mass functions of the GLIMPSE9 cluster from
[F222M] or [F160W] magnitudes, for  ages of 15, 20 and 27 Myr, are shown with a continuous line.  
The same function after correction for incompleteness is shown with a dashed line. Error bars represent 
the statistical Poisson uncertainty. Linear fits to bins with log(M/\Msun) from 0.1 to 0.85-0.90 
(i.e., F222M between 17.5 and 6.0 mag) are also shown.} 
\end{figure}

\begin{figure}[!] 
\begin{centering}
\resizebox{0.5\hsize}{!}{\includegraphics[angle=0]{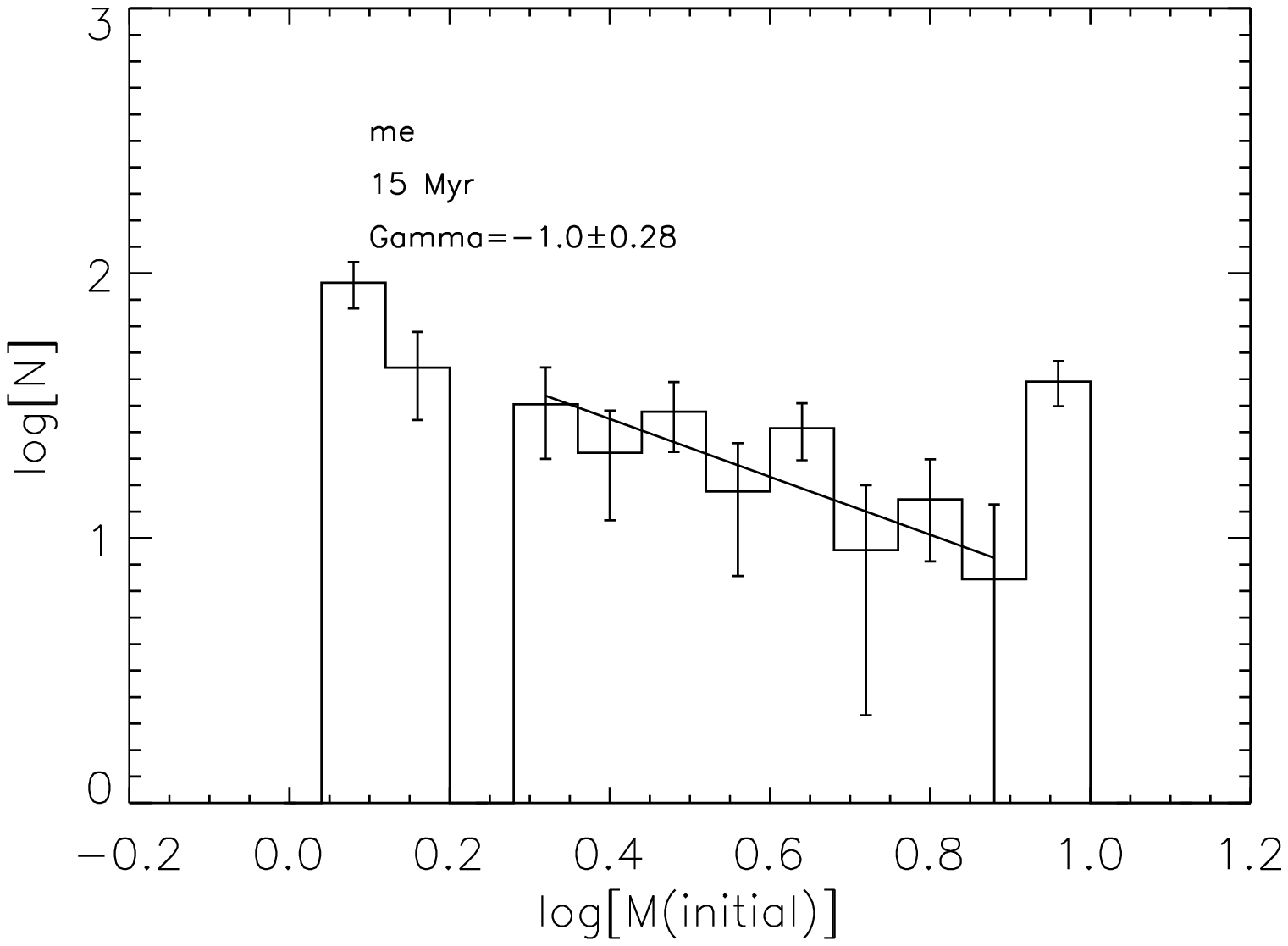}}
\resizebox{0.5\hsize}{!}{\includegraphics[angle=0]{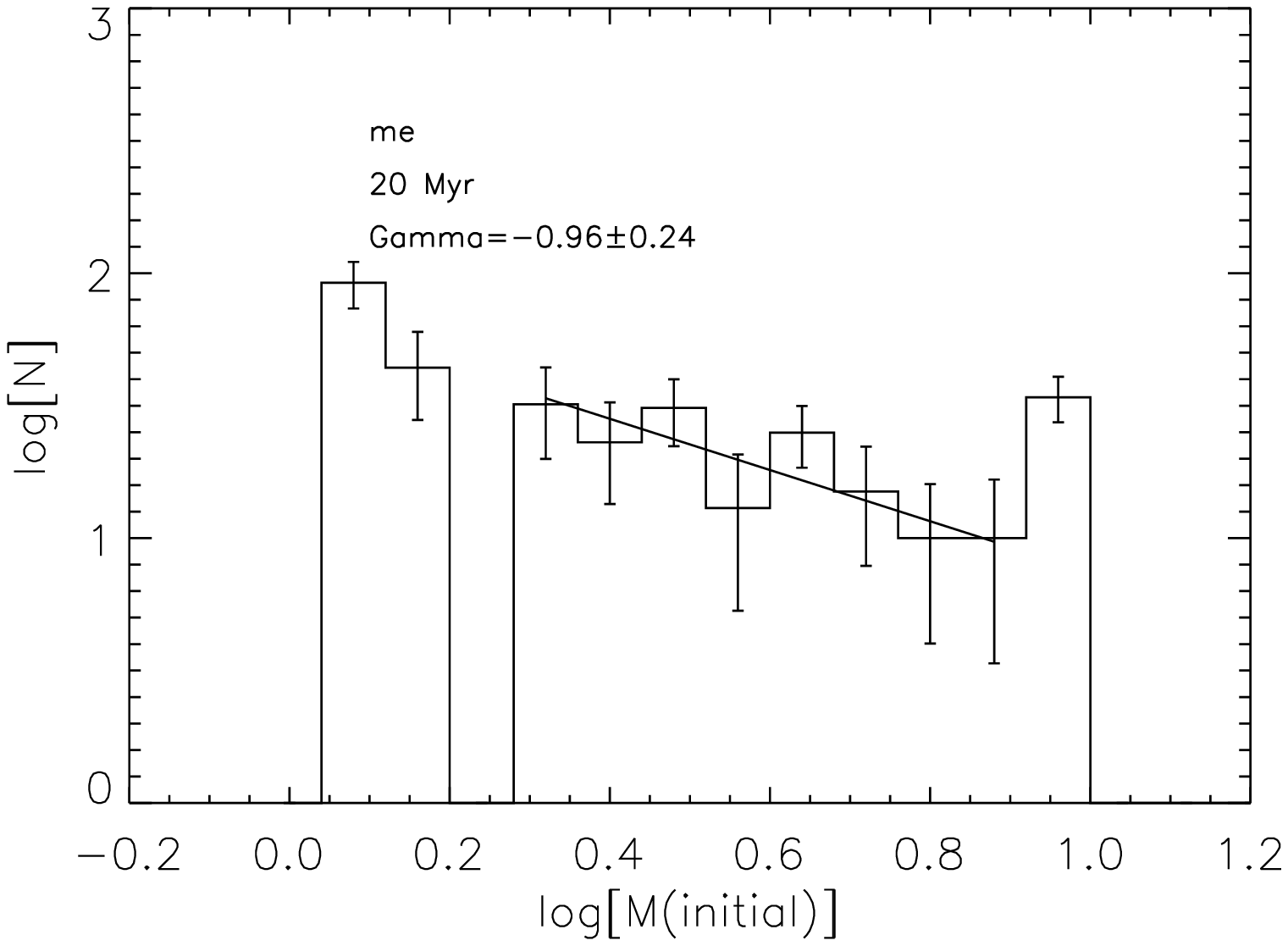}}
\resizebox{0.5\hsize}{!}{\includegraphics[angle=0]{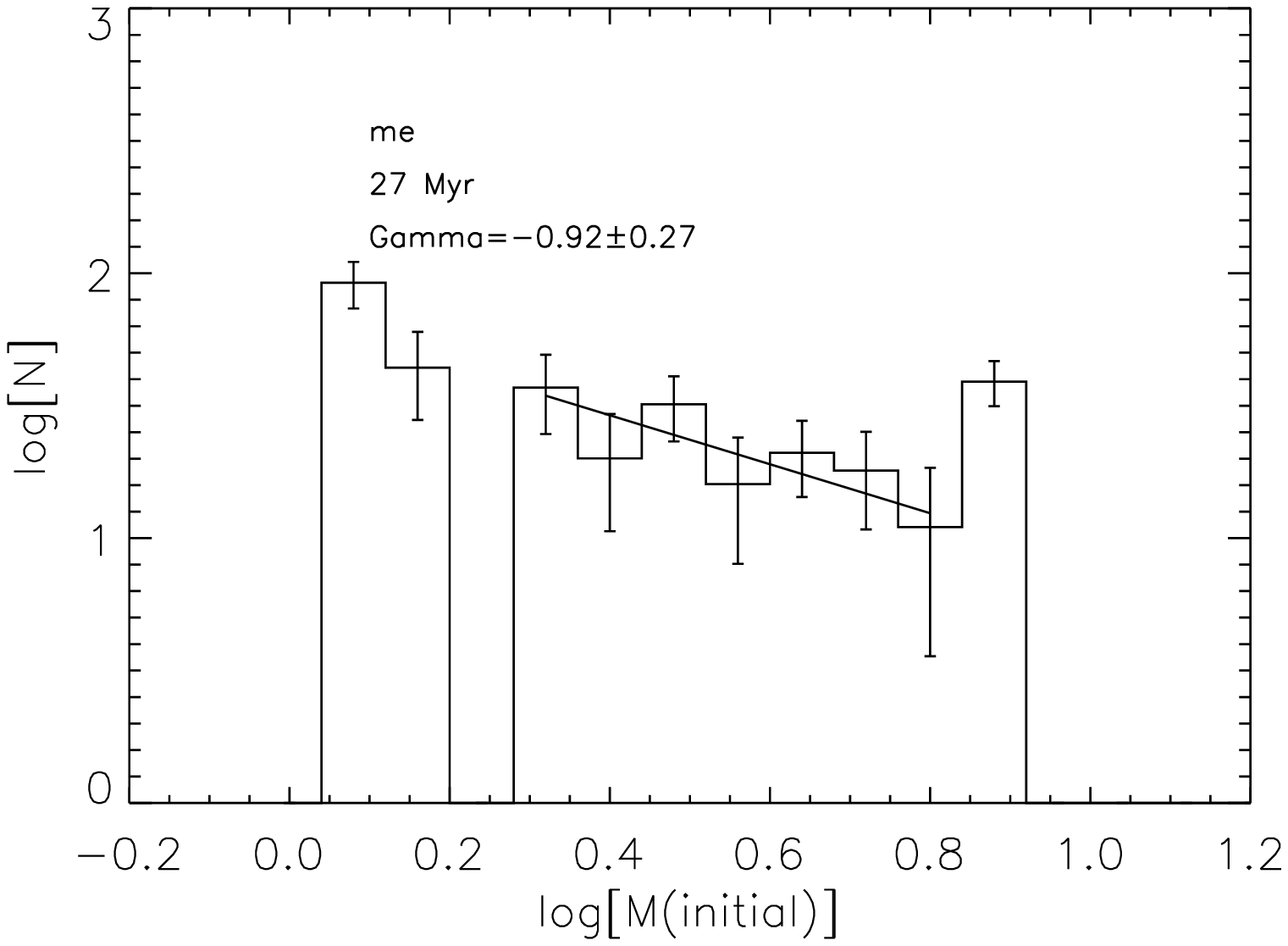}}
\end{centering} 
\caption{\label{gl09spectra2}  Background-subtracted mass functions of the GLIMPSE9 cluster from the
extinction corrected $m_e$ magnitudes. Symbols as in Fig.\ \ref{gl09spectra}}.
\end{figure}

Apparent magnitudes of cluster members can be transformed into
initial masses by assuming an interstellar extinction,  a distance, and using an isochrone. We
considered  \Aks\ $=1.6\pm0.2$ mag,  as estimated  from the CMD (see above), and a distance of
4.2 kpc \citep{leahy08}.  In addition, we used  non-rotational models from the Geneva group, 
with increased mass-loss, Solar abundance, ages of 15, 20, and 27 Myr \citep{meynet94}, and  
with the color  transformation for Johnson filters by \citet{lejeune01}. The relations between 
the actual mass and the apparent \K -band magnitude for a 15, 20 and 27 Myr populations are shown in Fig.\ \ref{masslum}. 
We used the equations by \citet{kim05} to transform the theoretical isochrones for the F160W, 
F222M, and \Ks\ short filters. An average difference of $0.07\pm0.05$ mag is found between the  \Ks\
magnitudes from \citet{kim05} and those with the empirical transformation given in Sect.\ 2.2.
This  is a measure of the transformation uncertainty.
Mass functions for the GLIMPSE9 cluster for different ages and bands are shown in Figs.\ 
\ref{gl09spectra} and \ref{gl09spectra2}. 

We built mass functions for stars detected in the F160W mosaic, as well as for stars in the
F222M mosaic.  We selected  a mass range log(M/\Msun) from 0.1 to 0.8-0.9
to ensure that stars were above a completeness limit of 80\% (see the Appendix), and to include only 
main-sequence stars (see Fig.\ \ref{masslum}). Evolved stars fall in a single bin because
of the degenerate mass-luminosity relation (Fig.\ \ref{masslum}).
For an age of 15 Myr, a linear fit to the  mass function  in the mass range from
log(M/\Msun) = 0.1 to 0.85  with a bin size of log(M/\Msun)=0.05 yielded a slope of $-0.77 \pm
0.20$ when using the F160W magnitudes,  and of $-0.70 \pm 0.25$ when using the F222M
magnitudes.   For an age of 20 Myr, the average slope from the two bands is 
$-0.63 \pm 0.05$, while for an age of 27 Myr is $-0.64 \pm 0.09$. 
Bin size variations from log(M/\Msun)= 0.035 to 0.07  yield slopes with a standard
deviation  within 0.1, i.e.  the uncertainty due to bin size variation is smaller than the
uncertainty of the fit. 

Besides this uncertainty due to age, the error in subtracting the foreground and background
population needs also to be considered.  If
the control field is not representative of the background and foreground population seen  towards
the cluster, a systematic error is introduced  when subtracting the field stars. The incorrect
field subtraction would propagate into the LF and mass distribution. As seen in Fig.\
\ref{hstfields}, the control field happens to be located in a dustier region.  From the CMDs, the
field appears to have an excess of red stars ($2< H-$\Ks $<  2.4$ mag), which implies  an extra
extinction of \Aks=$\sim0.5$ mag. When artificially reducing the attenuation of the control field
by \Ak$=-0.5$ mag, the slope of the initial mass function decreases by $\sim 0.33$ ($-1.1\pm0.1$,
$-1.0\pm0.16$, $-1.0\pm0.1$ for 15, 20, and 27 Myr).  We also calculated a mass function with the
extinction free magnitudes $m_e$, which takes into  account differential extinction, and measured
a slope of   $-1.0\pm 0.3$, $-0.96\pm 0.24$, and $-0.92\pm 0.27$ for 15, 20, and 27 Myr.

The origin and universality of the initial mass function remain under very active investigation and
discussion. Typically, it is assumed as a power law with  an exponent of $\Gamma=-1.35$
\citep{salpeter55}.  Studies of starburst clusters, such as the Arches cluster, report flatter
functions \citep[e.g. $-0.7\pm0.1$~][]{figer99}.  Recent re-determinations of the Arches MF slope by
\citet{espinoza09} indicate a slope of $-1.1\pm0.2$, which is closer to a Salpeter. 

The GLIMPSE9 cluster shows a  mass function slightly flatter than the Salpeter's one. We tried
several ways to measure the mass function,  including harshly de-reddening the control field,  and
yet we consistently come up with slopes that are shallower than Salpeter.  Mass segregation, which
bring massive stars to sink into the cluster center and low mass stars to disperse into the field,
seems a plausible explanation for this \citep{kim06}.  However, considered the uncertainties in age
and background subtraction,  and the fact that the GLIMPSE9  cluster is several million years old, 
a Salpeter initial  mass function cannot be excluded.

\subsection{Cluster mass}    
By integrating the masses of the candidate member stars  (using our background-subtracted mass function)
down to a  mass of 1.0 \Msun, we measured a cluster mass of $1500\pm300$ \Msun, where the error
takes into account the age uncertainty. Since systematic errors could be due to errors in the 
field subtraction, we also estimate the cluster mass by integrating only to \Ks = 15 mag, i.e., by using the
upper part of the diagram where field contamination is negligible, and extrapolating to 1.0 \Msun\
with a power law. When using the slope we estimated from the  mass distribution,  a mass of $1600
\pm 400$ \Msun\ is obtained, while with the Salpeter mass distribution the mass value rises to
$2100\pm 200$ \Msun.

\section{Cluster surroundings} 
\label{surrounding}

\subsection{A giant molecular cloud with SNRs}

{
\begin{deluxetable*}{llllrrl}
\tablewidth{0pt}
\tablecaption{\label{table:snr} Positions of the supernovae from literature.}
\tablehead{
\colhead{ID}& 
\colhead{NAME}& 
\colhead{RA(J2000)}& 
\colhead{DEC(J2000)} & 
\colhead{velocity(\kms)} &
\colhead{diameter(\arcmin)} &
\colhead{References} 
}
\startdata
1 & SNR/W41                         & 18 34 46.42 & $-$08 44 00 & 77$\pm$5 &30.0  & \citet{green91,leahy08}\\
\hline
2 & SNR22.7-0.2                     & 18 33 17.86 & $-$09 10 35 & \nodata   &30.0  & \citet{green91}\\         
  & G022.8$-$0.3                    & 18 33 45.50 & $-$09 09 17 & 82.5     &10.9  & \citet{kuchar97}\\
\hline
3 &  $G22.7583-0.4917$ & 18 34 28.30& $-$09 16 00 & 74.8     & 4.7  & \citet{kuchar97} \\
  &                                 & 18 34 28.00 & $-$09 16 00 & 76.0     &      & \citet{bronfman96}\\
  &			            & 18 34 26.70 & $-$09 15 50 & \nodata   & 5.0  & \#33 in \citet{helfand06}\\
\hline
4 &  $G22.9917-0.3583$              & 18 34 26.59 & $-$09 00 09 & \nodata   & 4.5  & \#34 in \citet{helfand06}\\
  &  $G22.9-0.3$                    & 18 34 12.60 & $-$09 01 20 &  70.9     & 4.8  &   \citet{kuchar97}\\      
\hline
5 &  $G23.5667-0.0333$              & 18 34 17.09 & $-$08 20 21 & \nodata   & 9.0  & \#35 in \citet{helfand06}\\
  &  $G23.5-0.0$                    & 18 34 19.60 & $-$08 22 17 & 91.3      & 6.1  & \citet{kuchar97}\\     
\enddata
\tablecomments{Positions and dimensions were measured on the MAGPIS image.} 
\end{deluxetable*}
}

\begin{figure}[!]
\begin{centering}
\resizebox{0.9\hsize}{!}{\includegraphics[angle=0]{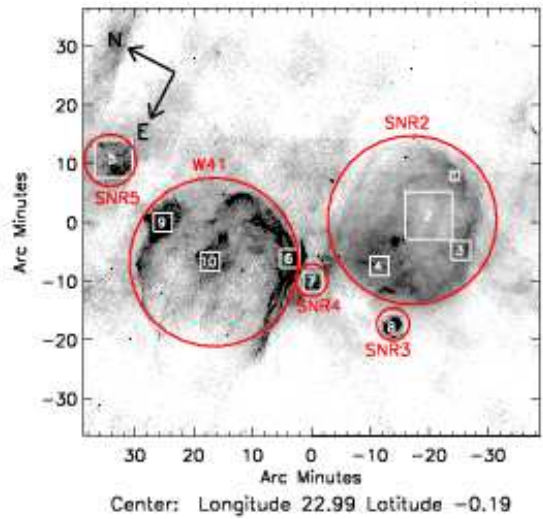}}
\end{centering}
\caption{\label{snr} 90 cm images from the Multi-Array Galactic Plane Imaging Survey (MAGPIS) 
\citep{white05}. Grey circles show the 
locations of the SNRs listed in Table \ref{table:snr}. White squares show the position 
of regions listed in Table \ref{table:spectralindex}, for which we calculated a 
spectral index. Longitude is along the x-axis, while latitude is along the y-axis.      
}
\end{figure}


The GLIMPSE9 cluster is located at (l,b)=(22.76 \degr, -0.40 \degr), in the
direction of a  GMC \citep{dame01}. The CO emission from
the molecular cloud peaks  near l=23.3 \degr, b=-0.3 \degr;  it extends over two
degrees in longitude with a line-of-sight velocity  from 70 to 85  \kms\ 
\citep{dame01}.  \citet{albert06} derived an upper limit for the  total H$_2$
mass  of 2.1 $\times 10^6$ \Msun, assuming a  (near) kinematic distance of 4.9
kpc. The linear size of about 100 pc is in the range of other known giant 
molecular clouds.
Several SNRs coincide in projection with this molecular cloud (see Fig.\ 
\ref{snr}). Two  SNRs are listed  in the catalog of \citet{green91}, 
G022.7$-$00.2   and  G023.8$-$00.3 (W41).   Three other candidate SNRs were
identified by \citet{helfand06}, G$22.7583-0.4917$, G$22.9917-0.3583$, and G$23.5667-0.0333$.   
\citet{leahy08} concluded that the SNR W41 is associated with the GMC, and  has
a radial velocity $v_{LSR}$ = 77.0 \kms.  \citet{kuchar97} 
report $v_{LSR}$ = 82.5 \kms\ for G022.8$-$00.3, and $v_{LSR}$ = 91.3 \kms\ for G23.5$-$0.0. 
\citet{bronfman96} measured a $v_{LSR}$ =
76.0 \kms\  towards G22.7583$-$0.4917. This suggests that at least three candidate
SNRs are associated with the same GMC. In Table \ref{table:snr} we report
positions of the  candidate SNRs and associated line-of-sight  velocity, when
available.  CO observations   confirm a molecular cloud with a peak
at $V_{LSR}=76\pm5$ \kms\ and a full-width-half-maximum of  22 \kms\ 
at the location of the GLIMPSE9 cluster  \citep{dame01}.

The classification of G022.7$-$00.2 is reported  as uncertain by
\citet{green91}, and the same radio source is listed as an HII region in other
works \citep[e.g.][]{kuchar97,paladini03}.  We used archival radio data at 20
and 90 cm from the MAGPIS \citep{white05}  to measure the radio spectral index 
of the candidate SNRs (see Fig.\ \ref{snr} and  Table
\ref{table:spectralindex}). Given the negative spectral indexes, the radio
emission appears dominated by synchrotron emission in all cases. Since
G$22.9917-0.3583$  and part of the radio shell  of G022.7$-$00.2 overlaps with
GLIMPSE 8 \um\  emission, they are probably a composition of SNRs  and HII
regions.   Assuming pure circular motion, a peak velocity of 78 \kms, a distance
of 7.6 kpc for  the Galactic Center, and a Solar V=214 \kms, a systematic
deviation of 5 \kms, as well  as a random deviation of 5 \kms, \citet{leahy08}
derived a distance d=3.9-4.5 kpc for SNR/W41. 
Using an average distance of 4.2 kpc, the angular sizes of  G022.7$-$00.2  and  
G023.3$-$00.3 (W41)  ( $\sim$27\arcmin) yield a linear size of $\sim$33 pc,   
while the angular sizes of G$22.7583-0.4917$, G$22.9917-0.3583$ and G$23.5667-0.0333$ (about 
5 \arcmin) yield a linear size of $\sim$6pc. These physical sizes are well within the 
range of other Galactic SNRs \citep{stupar07}.
Recently, \citet{brunthaler09} measured trigonometric parallaxes of 4.6 and 5.9 kpc with
two methanol maser sources found towards G23.01$-$0.41 and G23.44$-$0.18,
suggesting  different complexes arranged along the line of sight. 



\begin{deluxetable}{rllrl}
\tablewidth{0pt}
\tablecaption{\label{table:spectralindex} Spectral index from 90 and 20 cm MAGPIS data.}
\tablehead{
\colhead{Region}& 
\colhead{RA(J2000)}& 
\colhead{DEC(J2000)} & 
\colhead{diameter(\arcsec)} &
\colhead{Spectral index} 
}
\startdata
 1& 18:32:37.114&$-$09:13:32.15& 77&$-$0.82\\
 2& 18:33:09.794&$-$09:12:45.87&384&$-$1.23\\
 3& 18:33:20.950&$-$09:20:17.81&154&$-$0.98\\
 4& 18:33:56.335&$-$09:09:13.63&154&$-$0.78\\
 5& 18:34:15.082&$-$08:20:41.93&307&$-$0.35\\
 6& 18:34:20.244&$-$08:55:07.61&154&$-$1.44\\
 7& 18:34:26.988&$-$09:00:23.31&154&$-$0.36\\
 8& 18:34:29.686&$-$09:16:03.12&230&$-$0.21\\
 9& 18:34:38.220&$-$08:33:10.46&154&$-$0.72\\
10& 18:34:47.170&$-$08:43:24.00&154&$-$1.18\\
\enddata
\end{deluxetable}

\subsection{Stellar candidate clusters.}

\begin{figure}[!]
\begin{centering}
\resizebox{0.9\hsize}{!}{\includegraphics[angle=0]{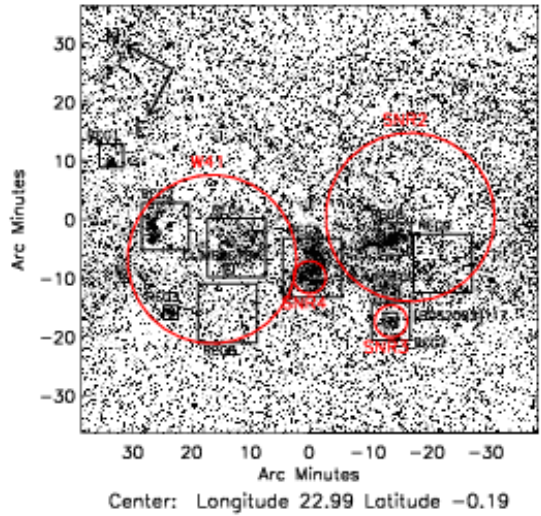}}
\end{centering}
\caption{\label{fig.clusters} The greyscale shows a 3.6 \um\ GLIMPSE image.  
Circles show the locations of the four SNRs.
Squares and labels show the regions in Table \ref{candclusters}.
Longitude is along the x-axis, while latitude is along the y-axis.
}
\end{figure}

The presence of three SNRs with similar velocities ($v_{LSR}$ =75-85 \kms)
suggests that massive star formation resulting in the production of several
massive O stars has been active in multiple sites of this GMC. The stellar
cluster GLIMPSE9 represents one episode of this star formation. Two other
candidate clusters are reported in literature in the direction of the same
molecular cloud. The $[BDS2003]$117 cluster \citep{bica03} appears to be
located  onto the SNR shell G22.7583$-$0.4917, and the  GLIMPSE10 cluster 
\citep[the candidate number 10 in the list by][]{mercer05} onto the SNR/W41 
(see Fig.\ \ref{fig.clusters}).

We searched for other stellar over-densities in the direction of this GMC. 
Detection of stellar over-densities are difficult due to the patchiness of the
interstellar extinction. With the exception of GLIMPSE9 cluster,  no clear
over-densities were detected in the 2MASS images. To overcome interstellar
extinction, we  built a density map of  point sources detected at 3.6 \um,
which resulted in many spurious clumps. When looking at the SPITZER/GLIMPSE
images, however, several  nebular emissions are seen in all four IRAC channels of the
SPITZER/GLIMPSE survey, indicating the presence of HII regions.  An increased
number of bright 3.6\um\ stars also appears associated with  some of these regions. We
therefore visually selected regions of nebular emission in all four IRAC
channels or apparent over-densities of bright stars at 3.6\um. 
A list of the selected regions is given in Table
\ref{candclusters}. 

We show 2MASS CMDs of these selected regions, together with  a comparison CMD 
of the GLIMPSE9 cluster, in Fig.\ \ref{2masscmd}.

Several stellar branches are seen in each CMD, and each branch appears broadened
by differential reddening.  A bluer sequence with $H$-\Ks $=\sim 0.4$ mag is
visible in all CMDs. A redder sequence with $H$-\Ks $=\sim 1$ mag, i.e., similar
to that of the GLIMPSE9 cluster, appears only at certain locations.  Assuming a
distance of 4.2 kpc for the GMC, and considering an average of 1.8 mag of visual
extinction per kpc, and the extinction law by \citet{messineo05}, a stellar
population associated with the GMC must have a minimum interstellar extinction
of \Aks=0.7 mag. The bluer  branch is probably due to a young stellar population
associated with a closer spiral arm. Based on the similarity of interstellar
extinction, we suggest that  branches seen at $H$-\Ks $=\sim 1$ mag are due to a
stellar population associated with the GMC, similar to the  GLIMPSE9 cluster. 

Region 1, in the direction of SNR G$23.5-0.0$, shows nebular emission, but its 2MASS
CMD does not show a stellar sequence with similar color as that of GLIMPSE9.

Region 2, in the direction of the SNR/W41, shows a peak of nebular emission in
all four IRAC channels, indicating an HII region. The 2MASS diagrams also show a
branch  of bright stars at $H$-\Ks $=\sim 0.8$ mag. 

Region 3 shows a concentration of bright stars; however, no clear sequences are
seen in the CMD.

Region 4 is also in the direction of SNR/W41. This region includes the GLIMPSE10
cluster, for which \citet{mercer05} gives a radius of 0.8\arcmin; GLIMPSE10 
coincides with a nebular peak emission, but the actual extension of the emission
region has a radius of about 5\arcmin. Several bright stars are detected in this
region, and the  2MASS CMD shows a  sequence at  $H$-\Ks $=\sim 1.0$ mag.

Region 5 coincides with the area covered by the SNR G22.9917$-$0.3583, as seen in
the 90 cm image.  A  sequence at  $H$-\Ks $=\sim 1.0$ mag is detected. 

Region 6 is a peak of a nebular emission that extends and connects to region 5.
The CMD lacks stars in the redder sequence. 

Region 7 coincides with the SNR G22.7583-0.4917. This region includes the
$[BDS2003]$117 cluster area by \citet{bica03}. 


Two other regions, region 8 and region 9, were randomly selected, as a comparison
fields. Region 8  does not show associated nebular emission, and is in the
direction of the SNR/W41. The CMD shows the blue sequence, but not a clear red
sequence. Region 9, which is at the outer edge of the complex,  does not show
nebular emission, and its CMD lacks a sequence at  $H$-\Ks $=\sim 1.0$ mag.

\begin{deluxetable*}{lrrrl}
\tablewidth{0pt}
\tablecaption{\label{candclusters} Cluster positions.}
\tablehead{
\colhead{ID}& 
\colhead{RA(J2000)}& 
\colhead{DEC(J2000)} & 
\colhead{Radius(\arcmin)} &
\colhead{References} 
}
\startdata
GLIMPSE9          & 18 34 09.59 & $-$09 13 53 &  0.3\tablenotemark{a}  & \citet{mercer05} \\  
$[BDS2003]$117    & 18 34 27.00 & $-$09 15 42 &  0.6\tablenotemark{b}  & \citet{bica03}   \\
GLIMPSE10         & 18 34 47.00	& $-$08 47 20 &  0.8\tablenotemark{b}  & \citet{mercer05} \\
Region1           & 18 34 15.08 & $-$08 20 42 &  1.2\tablenotemark{c}  &\\
Region2           & 18 34 41.09 & $-$08 34 22 &  4.0\tablenotemark{c}  & \\ 
Region3           & 18 35 32.22 & $-$08 41 56 &  1.2\tablenotemark{c}  & \\ 
Region4           & 18 34 31.59 & $-$08 46 47 &  5.0\tablenotemark{c}  & \\
Region5           & 18 34 20.00 & $-$08 59 48 &  5.0\tablenotemark{c}  & \\
Region6           & 18 33 36.03 & $-$09 10 01 &  2.7\tablenotemark{c}  & \\ 
Region7           & 18 34 27.69 & $-$09 15 52 &  3.3\tablenotemark{c}  & \\ 
Region8           & 18 35 14.46 & $-$08 50 34 &  5.0\tablenotemark{c}  & \\ 
Region9           & 18 33 35.88 & $-$09 19 08 &  5.0\tablenotemark{c}  & \\	 
\enddata
\tablecomments{ Coordinates are followed by radius and references. Region \#8 and \#9 are randomly selected. } 
\tablenotetext{a}{The radius is  measured in the \Ks-band image as the half light radius.}
\tablenotetext{b}{The radius is given in the referenced work.}
\tablenotetext{c}{The radius is  measured in the 3.6 \um-band image.}
\end{deluxetable*}


\begin{figure}[!]
\begin{centering}
\resizebox{0.5\hsize}{!}{\includegraphics[angle=0]{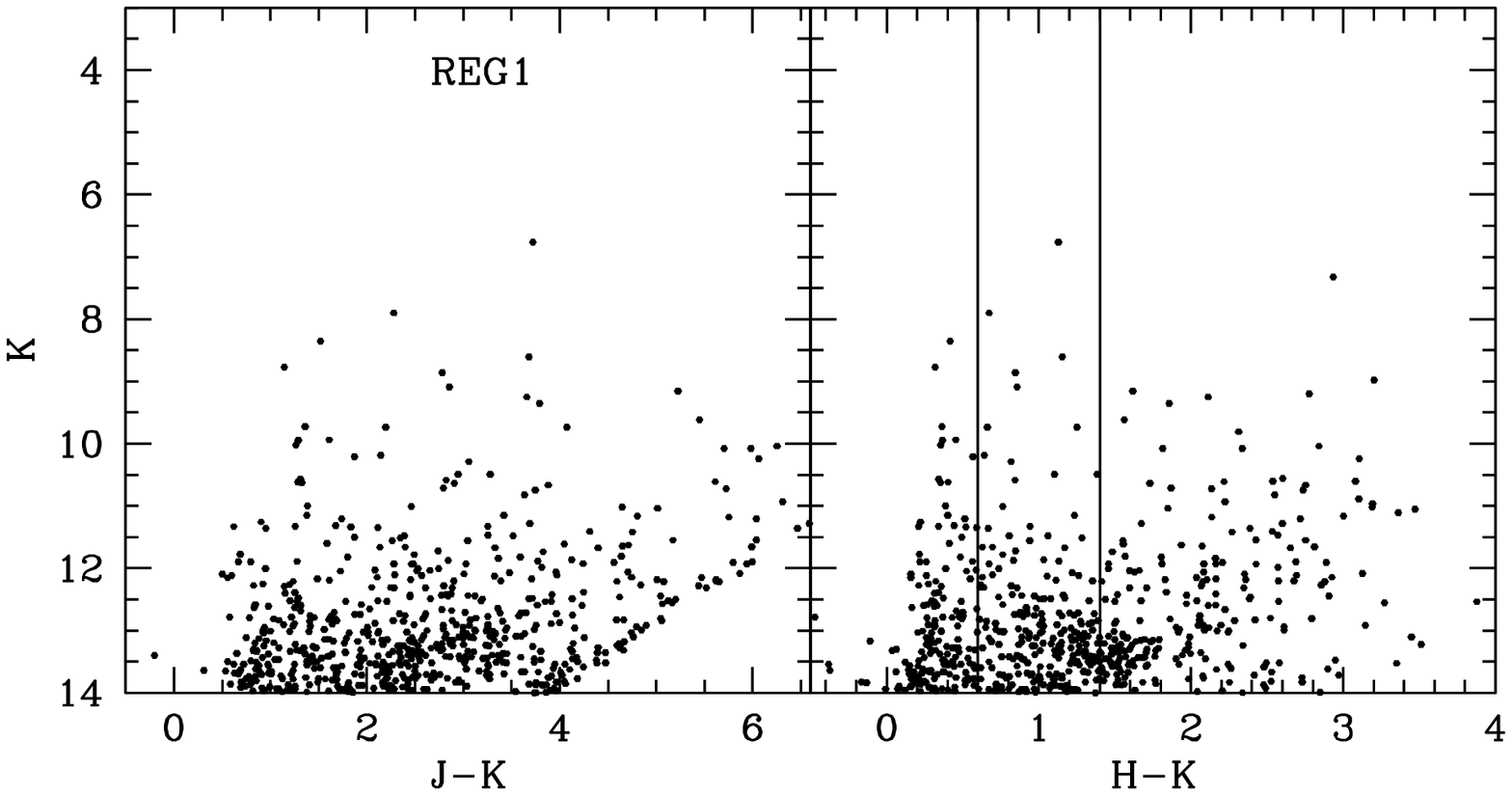}}
\resizebox{0.5\hsize}{!}{\includegraphics[angle=0]{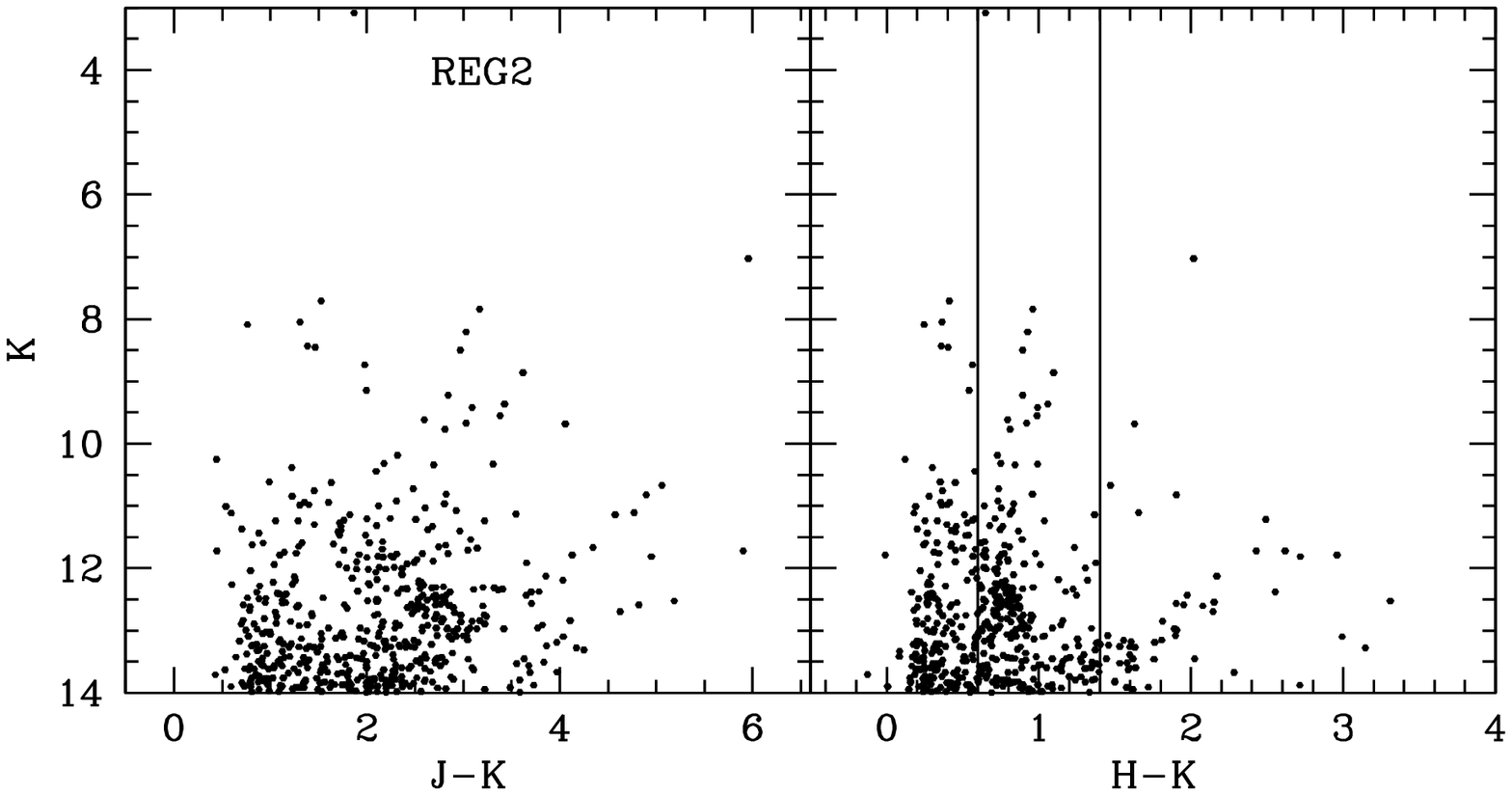}}
\resizebox{0.5\hsize}{!}{\includegraphics[angle=0]{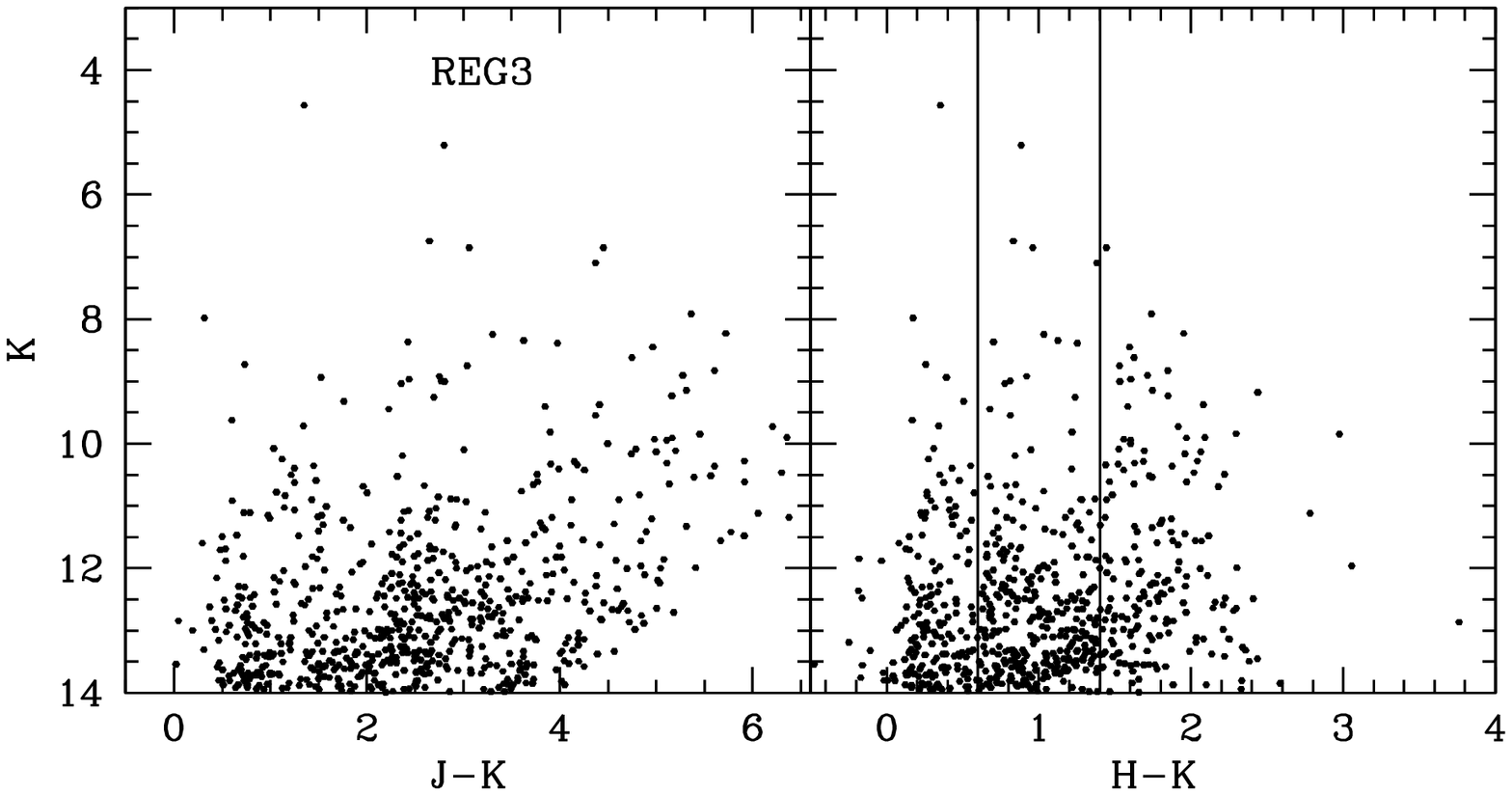}}
\resizebox{0.5\hsize}{!}{\includegraphics[angle=0]{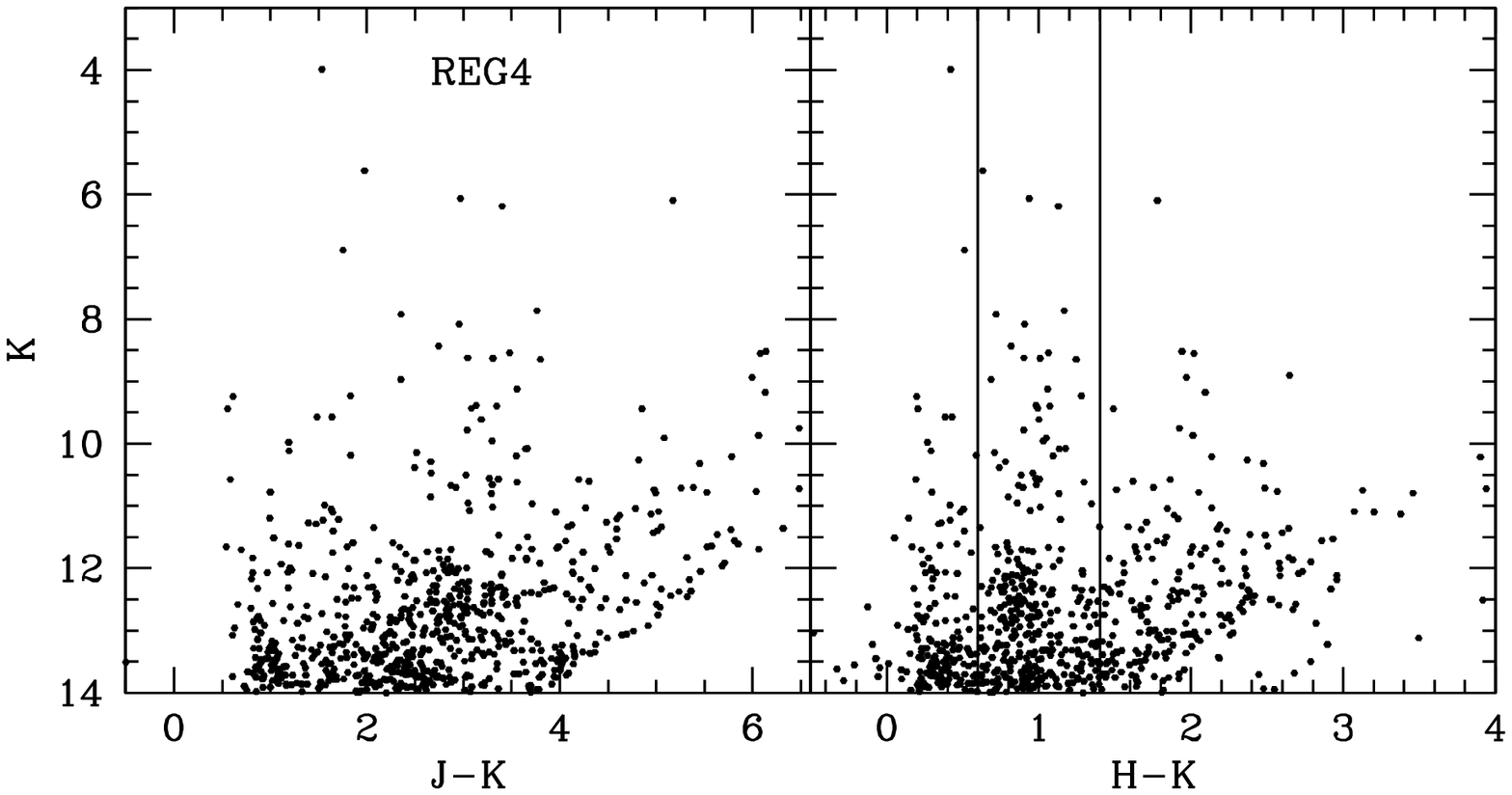}}
\resizebox{0.5\hsize}{!}{\includegraphics[angle=0]{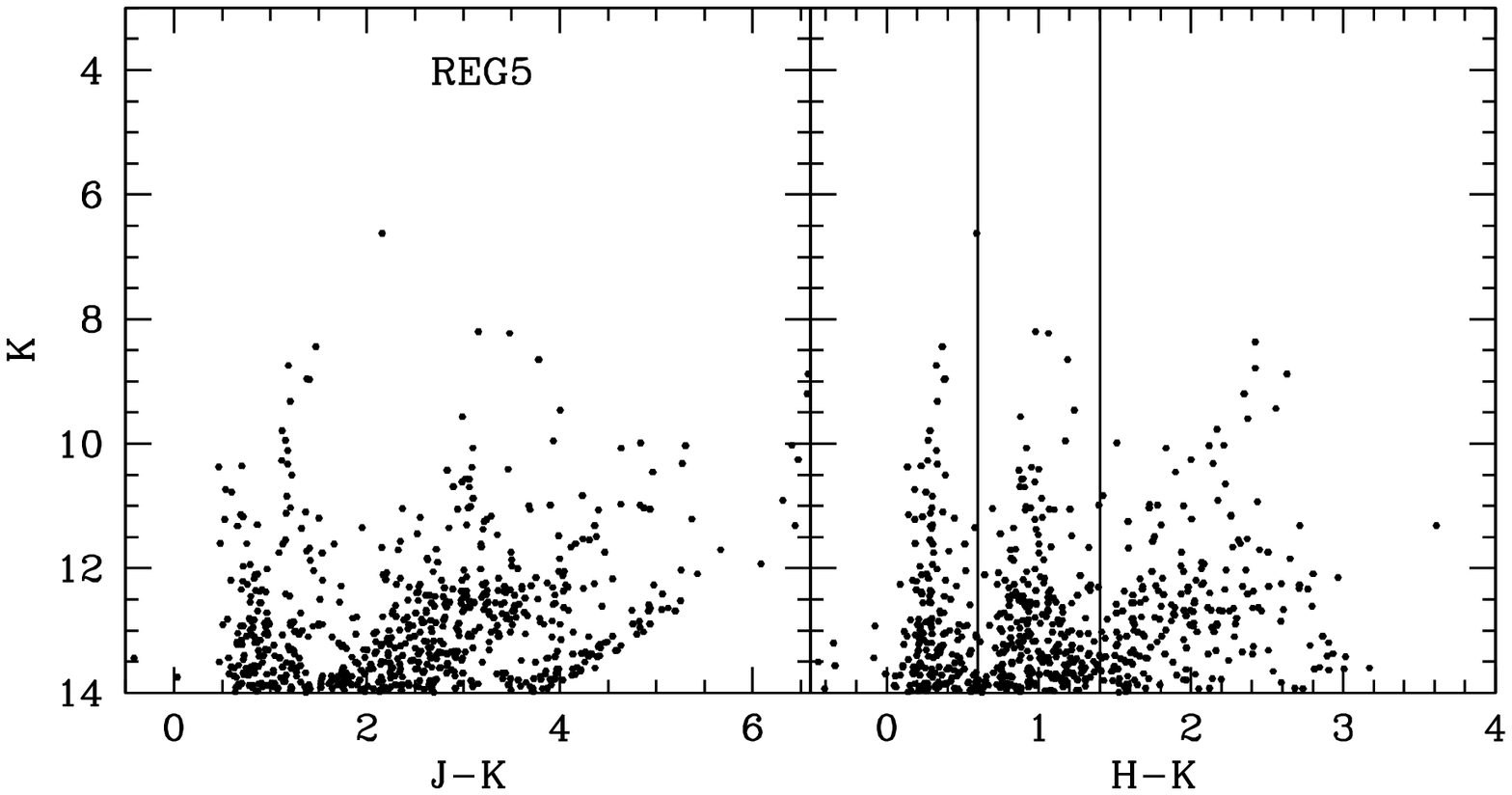}}
\resizebox{0.5\hsize}{!}{\includegraphics[angle=0]{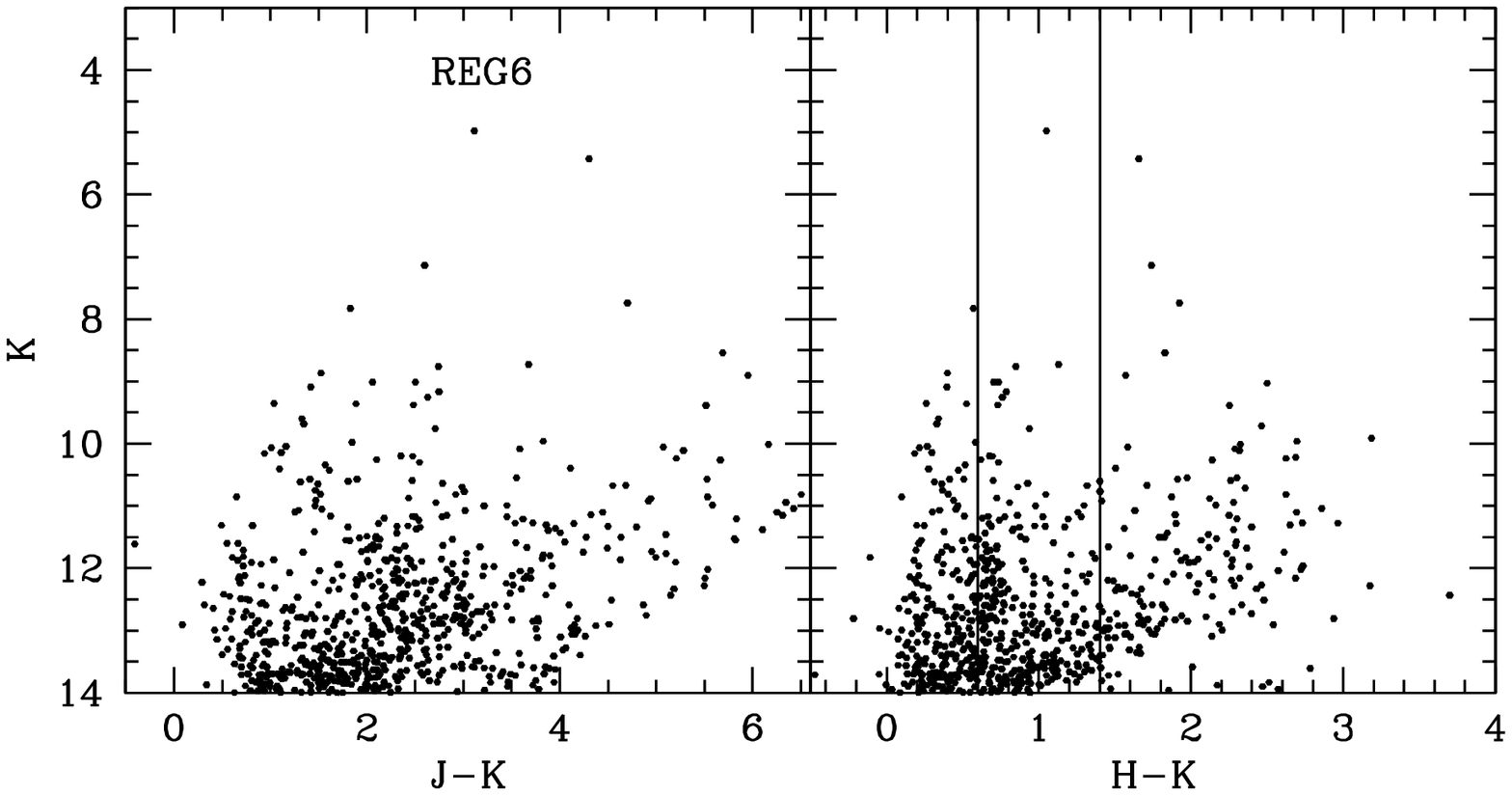}}
\end{centering}
\caption{\label{2masscmd} 2MASS J$-$\Ks\ versus \Ks, and H$-$\Ks\ 
versus \Ks\ diagrams of selected regions in direction of the
GMC (see Table \ref{candclusters}). For clarity,  an area of 3\arcmin\ is used
for all CMDs. Two vertical lines at
($H-$\Ks$=0.6$ and 1.4 mag indicate the color region of the GLIMPSE9 sequence.
A similar sequence is visible in reg \#2,\#4, \#5 and \#7.
}
\end{figure}

\begin{figure}[!]
\addtocounter{figure}{-1}
\begin{centering}
\resizebox{0.5\hsize}{!}{\includegraphics[angle=0]{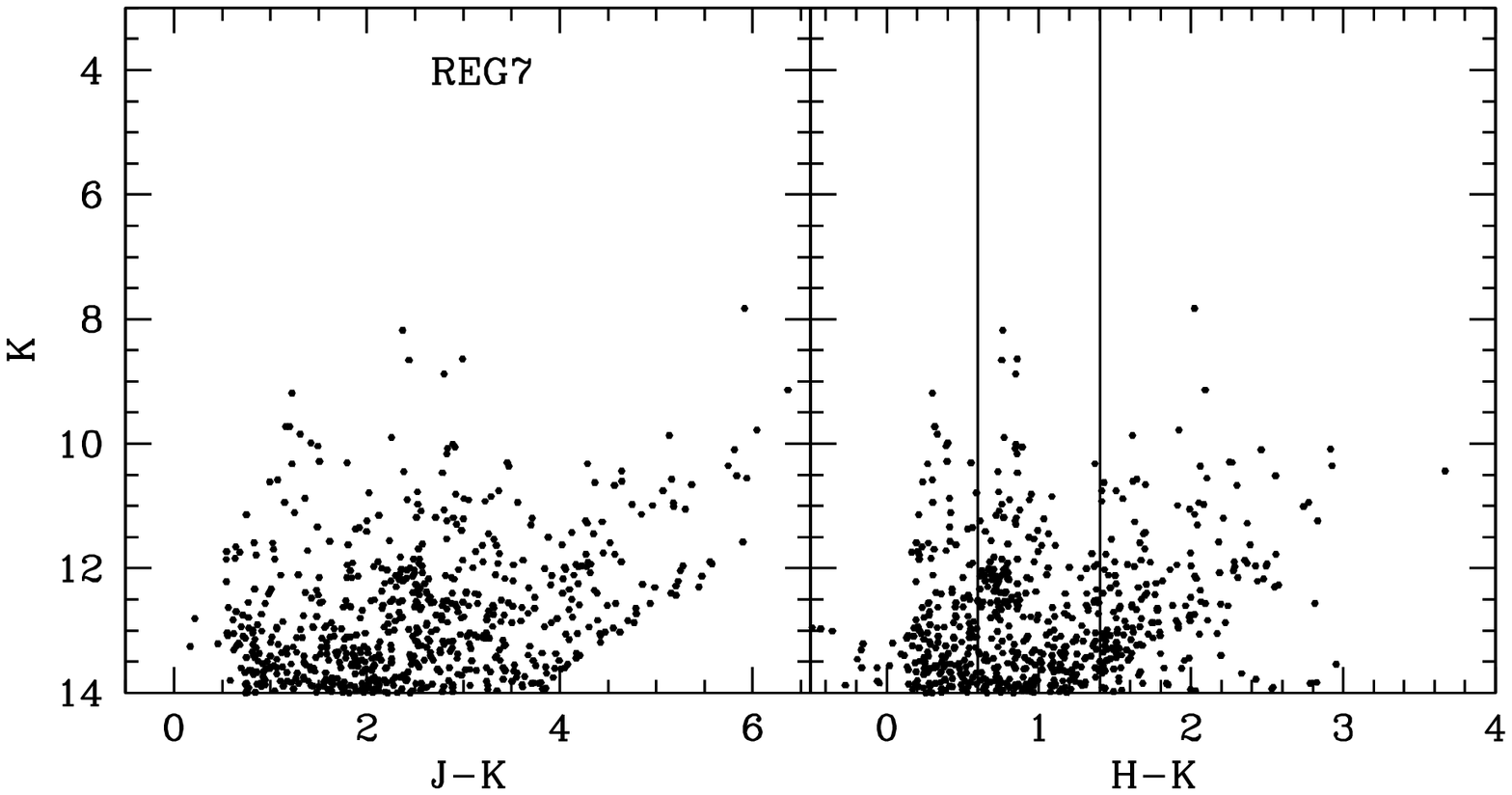}}
\resizebox{0.5\hsize}{!}{\includegraphics[angle=0]{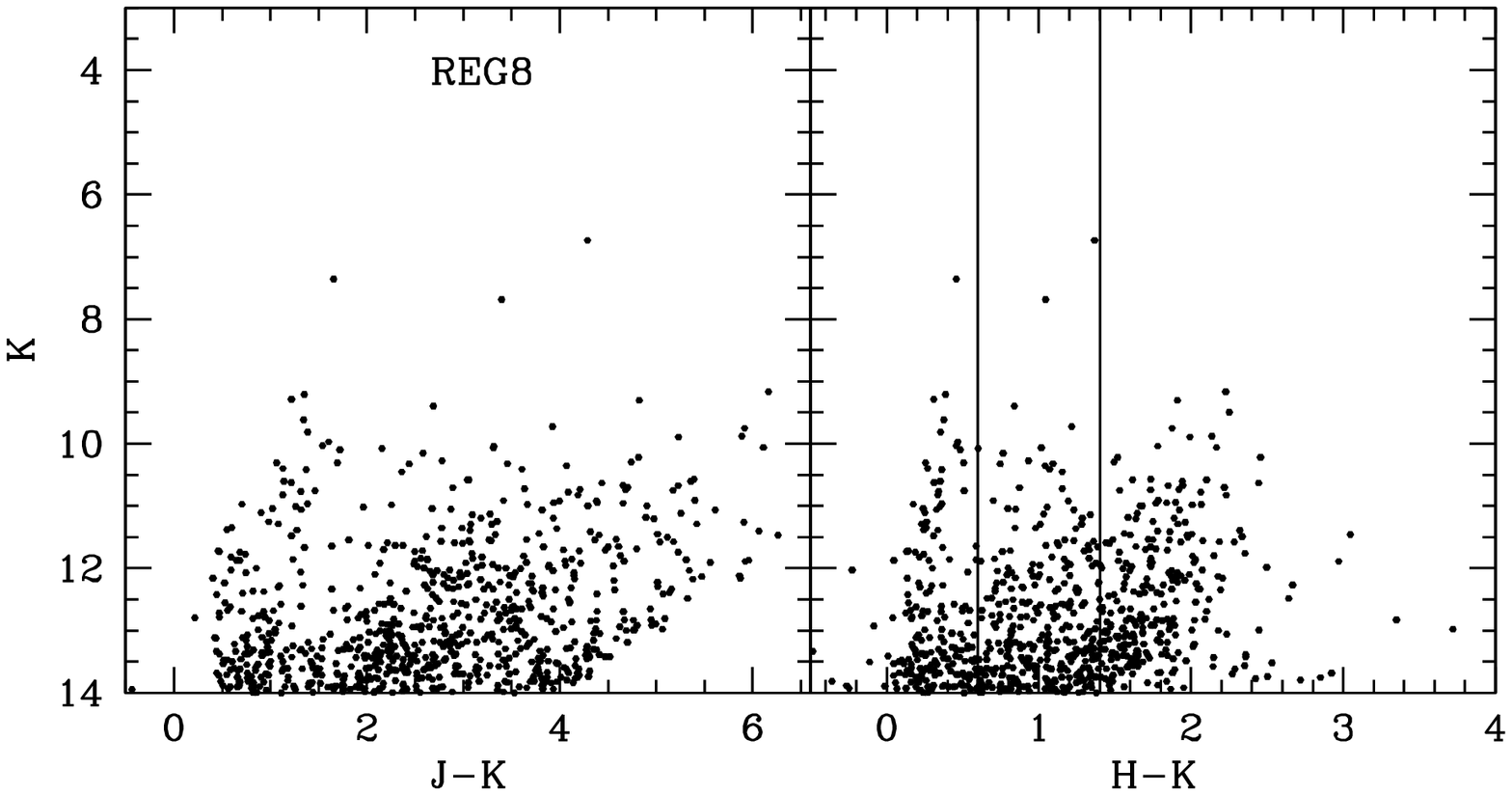}}
\resizebox{0.5\hsize}{!}{\includegraphics[angle=0]{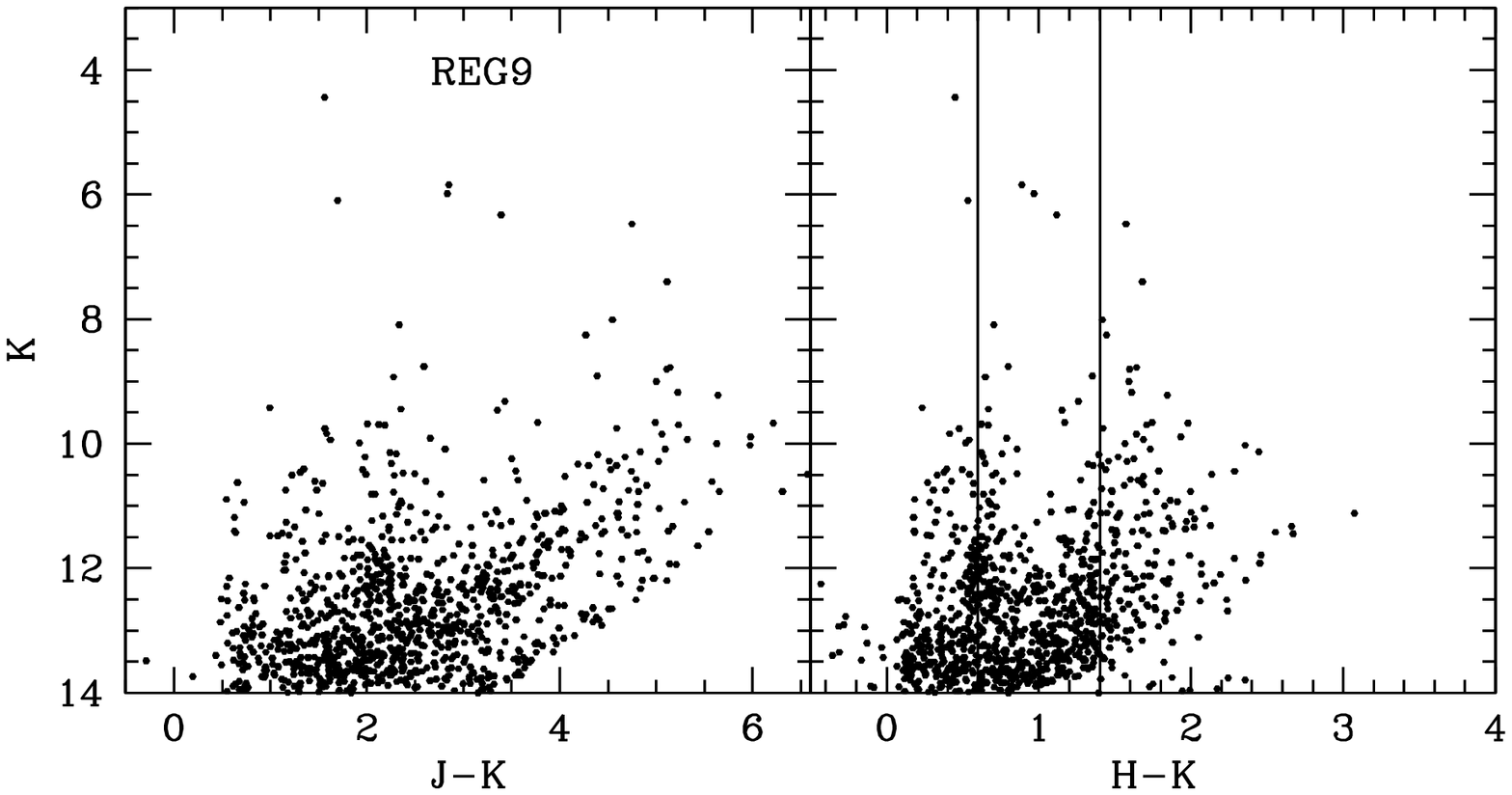}}
\resizebox{0.5\hsize}{!}{\includegraphics[angle=0]{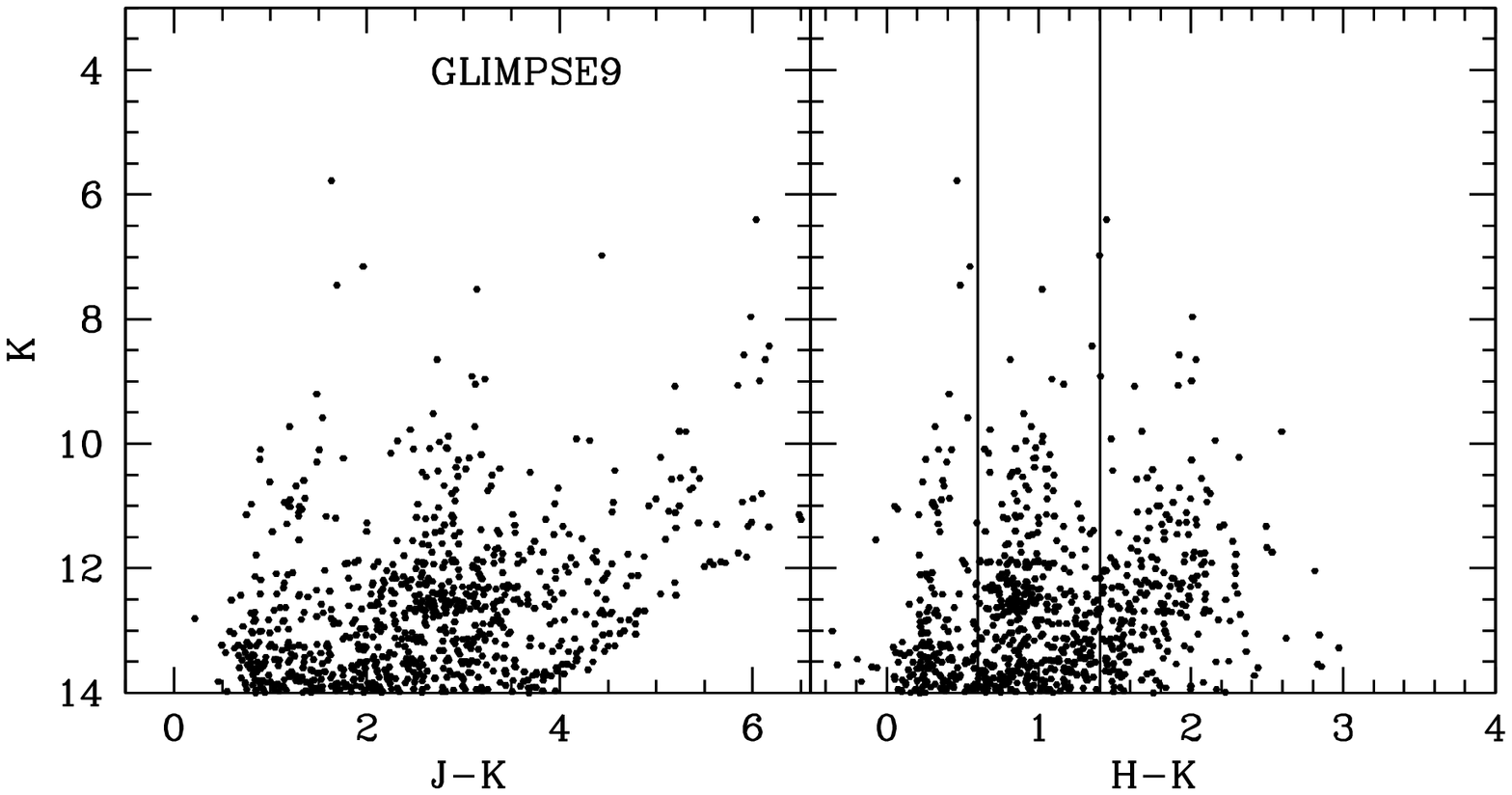}}
\end{centering}
\caption{\label{2masscmd2} \ref{2masscmd} continued.}
\end{figure}

\section{Summary} 
\label{summary}


Investigation with HST/NICMOS data confirms that object number 9 in the list by
\citet{mercer05} is a stellar cluster with a well defined sequence in the $H-$\Ks\  versus
\Ks\ diagram. Low-resolution spectroscopic observations in K-band yield the spectral
types of the brightest candidate members, and confirm the presence of massive stars. Three
RSGs are detected, two of which are candidate cluster members, and two BSGs. A
spectrophotometric distance of $4.2\pm0.4$ kpc is derived, an age  between 15 and 27 Myr,
and a mass of at least $1600\pm400$ \Msun. 
The cluster   mass function appears slightly flatter than the Salpeter's one,
which could be the effect of mass segregation.

The cluster is located in the direction of a molecular complex, which hosts several SNRs
and  HII regions. The cluster distance  agrees well with that inferred for the complex by
\citet{leahy08}. A stellar population possibly associated with the giant molecular
cloud is seen in four other regions: two regions  are associated with the SNR/W41, one
region  with the SNR G22.9917$-$0.3583, and another  region  with SNR G22.7583-0.4917.  The
detection of massive stars in GLIMPSE9, and the concomitant presence of several SNRs
render this GMC particular interesting. It is a good laboratory to
investigate  various issues about massive stars and  
multi-seeded star formation.  By identifying stellar clusters of the same complex, one can study  their properties and
variations across space and time. Some of the most massive stars of the Milky Way may be
hiding among the cluster members. The detection  of short lived massive stars (e.g.
Wolf-Rayets, Red Supergiants, Luminous Blue Variables) is important to understand their formation, evolution
and fate. By association with the SNRs, massive stars also yield the initial masses of  the
supernova progenitors.  A more detailed study of this complex will be presented in a future
paper.

\acknowledgments { 
The material in this work is supported by NASA under award  NNG 05-GC37G, through the Long--Term Space
Astrophysics program. RMR acknowledges support from grant AST-0709479 from the National Science
Foundation.    This research was performed in the Rochester Imaging Detector Laboratory with support
from a NYSTAR Faculty Development Program grant.  
Based on observations with the NASA/ESA Hubble Space Telescope (GO program 11545, P.I. Ben Davies), 
obtained at the  Space Telescope Science Institute, which is operated by the Association of Universities 
 for Research in Astronomy (AURA), Inc., under NASA contract NAS5-26555.
This publication makes use of data products from the Two Micron All Sky Survey, which is a joint project
of the University of Massachusetts and the Infrared Processing and Analysis Center/California Institute of
Technology, funded by the National Aeronautics and Space Administration and the NSF.  
Data presented herein were obtained at the W.M. Keck Observatory, which is operated as a scientific
partnership among the California Institute  of Technology, the University of California, and the National
Aeronautics  and Space Administration. The Observatory was made possible by the  generous financial
support of the W. M. Keck Foundation.
IRMOS is supported by NASA James Webb Space Telescope, NASA Goddard Space Flight Center, STScI DDRF, and
KPNO.
This work made use of data from theMulti-Array Galactic Plane Imaging Survey (MAGPIS;White et al. 2005;
Helfand et al. 2006), and the MAGPIS Web site, http:// third.ucllnl.org/gps/index.html.
This research has made use of Spitzer's GLIMPSE survey data, the Simbad 
and Vizier database. 
The authors thank Dr. Dame for providing CO observations, and Dr.
Sungsoo Kim for interesting discussion about the stellar  mass function.
We thank Dr. Simon Clark, Dr. Karl Menten, and Dr. Valentin Ivanov for useful
discussions on giant molecular clouds and multi-seeded star formation.}

\appendix{
\section{Artificial-star experiments} 
For a proper cluster analysis (e.g. for a statistical  field decontamination, luminosity
function, and mass function studies),  one needs to  characterize all undesired 
biases of   photometric measurements in dense stellar fields.  Crowding limits the detection
of point sources, and such incompleteness varies from field to field because it depends on
the stellar density. We, therefore, run  simulations with artificial stars.

Artificial stars were created with the observed PSF.  Then,  the artificial stars were  
added into each mosaic at random positions, but imposing only one star over a $50 \times
50$ pixels area in order not to alter the level of crowding.   The luminosity function of
the artificial stars was assumed identical to the observed luminosity function of each
frame. The photometry of the artificial stars was recovered following the exact steps as
those performed for the actual catalog. The procedure was iterated 300 times, giving a 
total of 27,000 artificial stars per mosaic.  By comparing the fluxes of the artificial
stars' input with their fluxes after re-extraction, we obtained an estimate of the
photometric uncertainty and completeness limit.

In Figure \ref{complet} the differences between the input ($mag_{input}$) and  the extracted
magnitudes ($mag_{output}$) of the  artificial stars are shown as a function of
$mag_{input}$; differences remain, on average, null till a certain magnitude, then increase
with  increasing $mag_{input}$;  this is due to  blending between  artificial stars and real
stars, and it causes a bin-to-bin migration in the luminosity function (LF).   A star is
considered lost if  no star is located by the DAOPHOT point-source finding algorithm within
1.5 pixels (0.1\arcsec) of the inserted location.   The fraction of  recovered   stars per
bin of magnitude is  the completeness factor (\cf). A \cf\ above 80\%  is reached for 
$[F222M]<17.0$ mag in the cluster field,  and  $[F222M]<18.0$ mag in the control field. For the
cluster field,  the one sigma deviation is  below 0.3 mag for $[F222M] < \sim 16.0$
mag,   but 1.0 mag for   $[F222M] = \sim 18.5$  mag. In the control field,   the one sigma
deviation is below 0.3 mag for $[F222M] < 18$ mag,   and 1.0 mag for  $[F222M] = \sim
20$  mag.

\begin{figure}[!] 
\begin{center}
\resizebox{0.6\hsize}{!}{\includegraphics[angle=0]{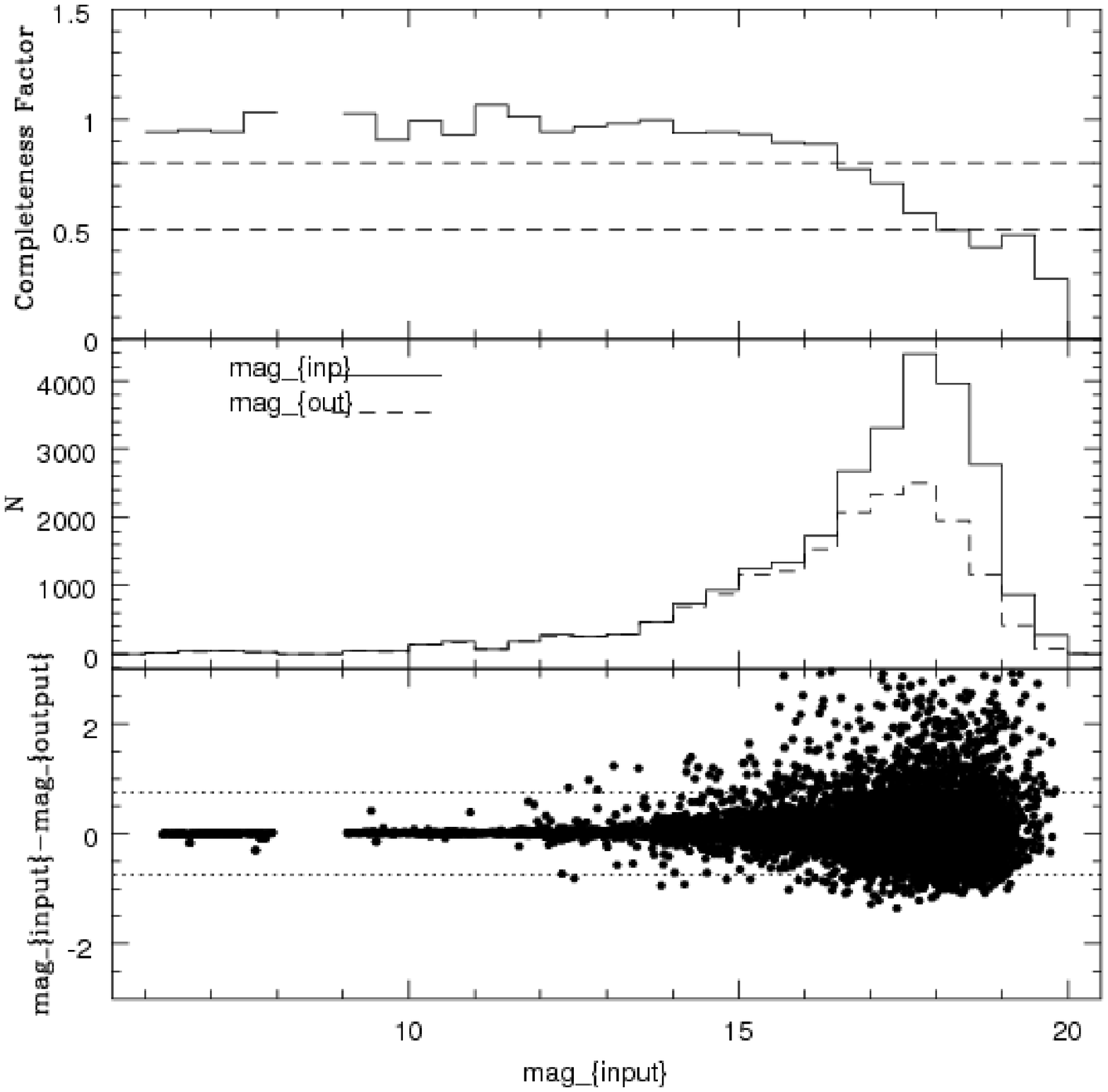}}
\end{center} 
\begin{center}
\resizebox{0.6\hsize}{!}{\includegraphics[angle=0]{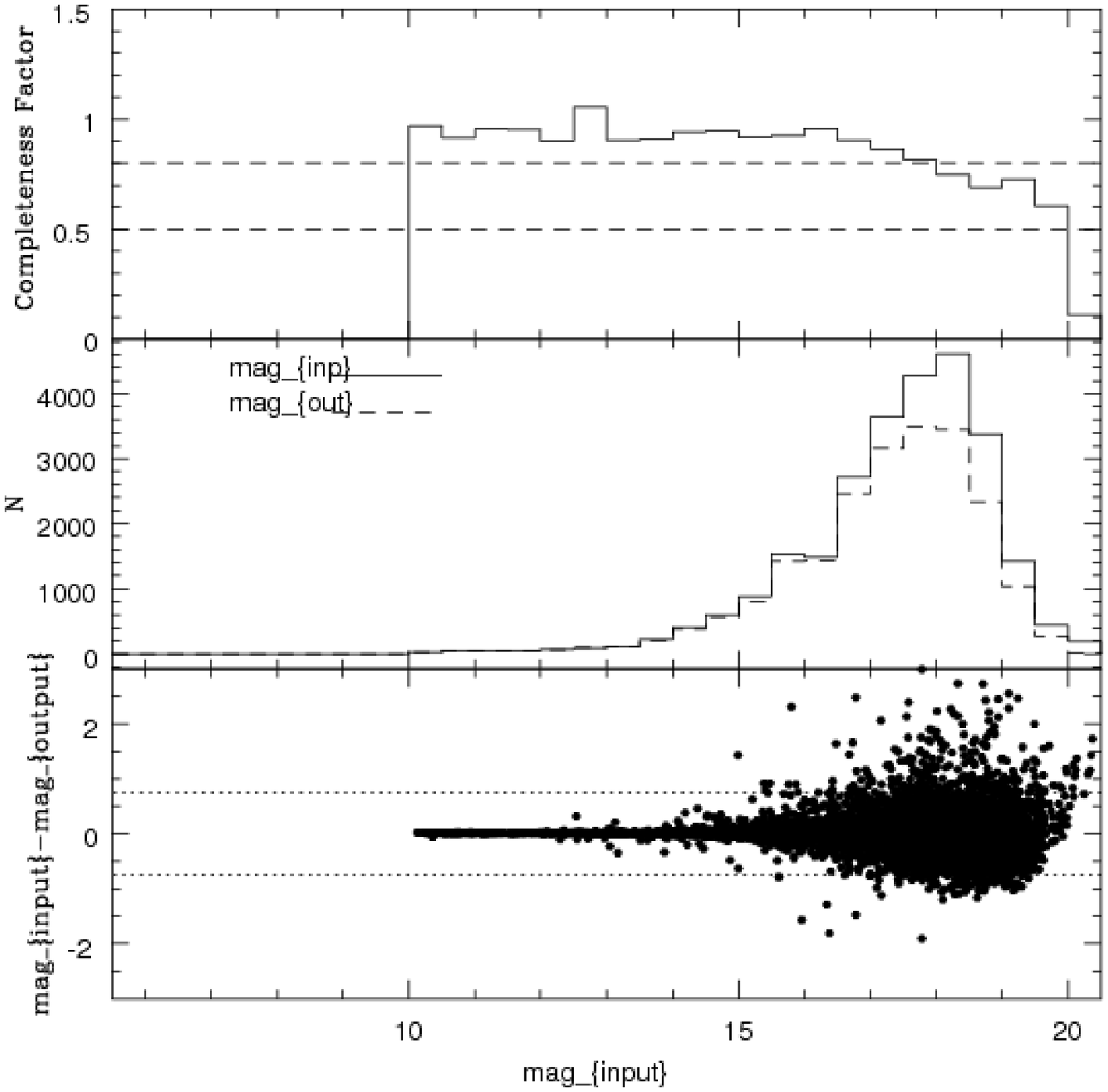}}
\end{center} 
\caption{\label{complet}  
Completeness experiment for the cluster field observation with the F222M filter.
Cluster data are shown in the top figure, and field data are shown in the
figure at the bottom. 
{\bf Lower-panel:} The differences between input and output  magnitudes are
shown ($mag_{input}-mag_{output}$). The dashed lines indicate a difference of
$\pm0.75$ mag. {\bf Middle panel} shows the distribution in magnitudes of
the artificial stars (a solid line is used for the input magnitudes, a
dashed line for the output magnitudes). {\bf Upper panel:}  The ratio between
the number of  stars  in output  and  in input is shown, as a function of 
magnitude. Two dotted lines indicated the 80\% and 50\% completeness limits.}  
\end{figure}

}

\end{document}